\RequirePackage{lineno}
\documentclass[twocolumn,aps,prc,showpacs,superscriptaddress,floatfix]{revtex4-2}
\usepackage{rotating}
\usepackage{epsfig}
\usepackage{hyperref}
\usepackage{color}
\usepackage{lineno}
\usepackage{url}
\usepackage{multirow}

\usepackage{amsmath}
\usepackage{graphicx}
\usepackage[caption=false]{subfig}

\graphicspath{ {./plot/}, {./image/} }
\DeclareGraphicsExtensions{ .pdf, .png}

\makeatletter
\renewcommand{\p@subsection}{}
\renewcommand{\p@subsubsection}{}
\makeatother

\makeatletter
\let\LN@equation\equation
\let\LN@endequation\endequation
\renewcommand{\equation}{\linenomath\LN@equation}
\renewcommand{\endequation}{\LN@endequation\endlinenomath}
\let\LN@gather\gather
\let\LN@endgather\endgather
\renewcommand{\gather}{\linenomath\LN@gather}
\renewcommand{\endgather}{\LN@endgather\endlinenomath}
\makeatother

\begin{document}


\newcommand{\der}{\text{d}}
\newcommand{\rp}{\Psi_{\text{RP}}}
\newcommand{\kal}{\xi}
\newcommand{\skal}{\xi^{*}}
\newcommand{\mult}{N_{ch}}
\newcommand{\npoi}{N_{POI}}
\newcommand{\refm}{\der N / \der \eta}
\newcommand{\vvese}{v_{2}}
\newcommand{\realpara}{\sigma_{\uparrow}}
\newcommand{\realperp}{\sigma_{\perp\uparrow}}
\newcommand{\shffpara}{\sigma_{\downarrow}}
\newcommand{\shffperp}{\sigma_{\perp\downarrow}}
\newcommand{\np}{n_{\pi}}
\newcommand{\npp}{n_{\pi^{+}}}
\newcommand{\npn}{n_{\pi^{-}}}
\newcommand{\nr}{n_{\rho}}
\newcommand{\vvp}{v_{2,\pi}}
\newcommand{\vvr}{v_{2,\rho}}
\newcommand{\rk}{K}


\title{Decipher the $R_{\Psi_{m}}$ correlator in search for the chiral magnetic effect in relativistic heavy ion collisions}

\author{Yicheng Feng}
\email{feng216@purdue.edu}
\address{Department of Physics and Astronomy, Purdue University, West Lafayette, IN 47907, USA}

\author{Jie Zhao}
\email{zhao656@purdue.edu}
\address{Department of Physics and Astronomy, Purdue University, West Lafayette, IN 47907, USA}

\author{Hao-jie Xu}
\email{haojiexu@zjhu.edu.cn}
\address{School of Science, Huzhou University, Huzhou, Zhejiang 313000, China}

\author{Fuqiang Wang}
\email{fqwang@purdue.edu}
\address{Department of Physics and Astronomy, Purdue University, West Lafayette, IN 47907, USA}
\address{School of Science, Huzhou University, Huzhou, Zhejiang 313000, China}

\date{\today} 


\begin{abstract}

\begin{description}
	\item[Background]
	The chiral magnetic effect (CME) is extensively studied in heavy-ion collisions at RHIC and the LHC.
	An azimuthal correlator called $R_{\Psi_{m}}$ was proposed to measure the CME.
	By observing the same $R_{\Psi_{2}}$ and $R_{\Psi_{3}}$ (convex) distributions 
	from A Multi-Phase Transport (AMPT) model,
	by contrasting data and model as well as large and small systems 
	and by event shape engineering (ESE), 
	a recent preprint (arXiv:2006.04251v1) from STAR suggests that 
	the $R_{\Psi_{m}}$ observable is sensitive to the CME signal and relatively insensitive to backgrounds,
	and their Au+Au data are inconsistent with known background contributions.
	\item[Purpose]
	We examine those claims by studying the robustness of the $R_{\Psi_{m}}$ observable 
	using AMPT as well as toy model simulations.
	We compare $R_{\Psi_{m}}$ to the more widely used $\Delta\gamma$ azimuthal correlator to identify their commonalities and differences.
	\item[Methods]
	We use AMPT 
	to simulate Au+Au, p+Au, and d+Au collisions at $\sqrt{s_{NN}} = 200 \text{ GeV}$,
	and study the responses of $R_{\Psi_{m}}$ to anisotropic flow backgrounds in the model.
	We also use a toy model to simulate resonance flow background and input CME signal to investigate	their effects in $R_{\Psi_{2}}$.
	Additionally we use the toy model to perform an ESE analysis to compare to STAR data
	as well as predict the degree of sensitivity of $R_{\Psi_{2}}$ 
	to isobar collisions with the event statistics taken at RHIC.
	\item[Results] 
	Our AMPT results show that the $R_{\Psi_{2}}$ in Au+Au collisions 
	is concave and apparently different from $R_{\Psi_{3}}$, in contradiction to the findings in STAR's preprint,
	while the $R_{\Psi_{2}}$ in p+Au and d+Au collisions are slightly concave.
	Our toy model ESE analysis indicates that 
	the $R_{\Psi_{2}}$ is sensitive to the event-by-event anisotropy $q_{2}$ 
	as well as the elliptic flow parameter $v_{2}$.
	The toy model results further show that $R_{\Psi_{2}}$ 
	depends on both the CME signal and the flow backgrounds,
	similar to the $\Delta\gamma$ observable.
	It is found that the $R_{\Psi_{2}}$ and $\Delta\gamma$ observables show 
	similar sensitivities and centrality dependences in isobar collisions.
	\item[Conclusions] 
	Our AMPT results contradict those from a recent preprint by STAR. 
	Our toy model simulations demonstrate that 
	$R_{\Psi_{2}}$ is sensitive to both the CME signal and physics backgrounds. 
	Toy model simulations of isobar collisions show 
	similar centrality dependence and magnitudes for the relative $R_{\Psi_{2}}$ strengths 
	as well as the relative $\Delta\gamma$ strengths. 
	We conclude that $R_{\Psi_{2}}$ and the inclusive $\Delta\gamma$ are essentially the same.
\end{description}

\end{abstract}



\maketitle


\section{Introduction}

In quantum chromodynamics (QCD),
topological charge fluctuations in vacuum can cause chiral anomality in local domains~\cite{Lee:1974ma,Morley:1983wr,Kharzeev:1998kz,Kharzeev:2007jp}.
Such domains violate the parity ($\mathcal{P}$) and charge-parity ($\mathcal{CP}$) symmetry.
If a strong enough external magnetic field is also present,
quark spins would be locked depending on their charge, either parallel or anti-parallel to the magnetic field.
As a result, charge separation along the magnetic field would emerge in those chirality imbalanced domains, 
which has observational consequences in the final state. 
This is called the chiral magnetic effect (CME)~\cite{Kharzeev:1998kz,Kharzeev:2007jp}.

In non-central heavy ion collisions, 
excited QCD vacuum is formed in the central collision zone,
whereas the spectator protons can provide an intense, transient magnetic field~\cite{Kharzeev:2007jp}.
Thus, the CME is expected to emerge in those collisions,
which, if observed, would be a strong evidence 
for local $\mathcal{P}$ and $\mathcal{CP}$ violation in the strong interaction.

The magnetic field created in heavy-ion collisions is, on average, 
perpendicular to the reaction plane 
(RP, spanned by the impact parameter and the beam direction). A RP-dependent charge correlation observable $\Delta\gamma$ has been proposed~\cite{Voloshin:2004vk} and widely studied at the Relativistic Heavy Ion Collider (RHIC)~\cite{Abelev:2009ac,Abelev:2009ad,Adamczyk:2014mzf,Adamczyk:2013hsi, Zhao:2017wck,Zhao:2017ckp}
and the Large Hadron Collider (LHC)~\cite{Abelev:2012pa, Khachatryan:2016got, Sirunyan:2017quh,Acharya:2017fau, Acharya:2020rlz}.
An alternate correlator, called $R_{\Psi_{m}}$ ($m=2$ or 3 is the azimuthal harmonic order), was also proposed~\cite{Ajitanand:2010rc, Magdy:2017yje}.
The premise was that the physics backgrounds should result in a convex $R_{\Psi_{2}}$ distribution 
and the CME signal should give a concave one. 
This was contradicted by other background studies~\cite{Bozek:2018aad}, including one by us~\cite{Feng:2018so}.

Recently, the STAR collaboration released results~\cite{Magdy:2020csm} using a modified $R_{\Psi_{2}}$ variable 
(see Sec.~\ref{RmDefinition}).
Their AMPT (a multiphase transport~\cite{Lin:2004en}) and 
AVFD (Anomalous Viscous Fluid Dynamics~\cite{Jiang:2018cpc, Shi:2017cpu}) model studies, 
suggest that $R_{\Psi_{2}}$ is sensitive to the CME signal and relatively insensitive to backgrounds. 
It is found that the AMPT $R_{\Psi_{m}}$ results are 
convex and equal between $R_{\Psi_{2}}$ and $R_{\Psi_{3}}$;
that the $R_{\Psi_{2}}$ in Au+Au collisions is concave 
and in p+Au and d+Au collisions are flat or convex;
and that the $R_{\Psi_{2}}$ distribution in an event shape engineering (ESE)~\cite{Schukraft:2012ah} analysis
is insensitive to the event-by-event anisotropy parameter 
which is in turn sensitive to the flow anisotropy. 
These findings led to the conclusion that the Au+Au data indicate a strong signal consistent with the CME 
that cannot be explained by known backgrounds.

Since the qualitative features of the AMPT results by STAR~\cite{Magdy:2020csm} contradict 
the other similar background studies~\cite{Bozek:2018aad, Feng:2018so},
further investigations are warranted. 
In this paper, we first revisit our earlier AMPT study using the modified $R_{\Psi_{m}}$ variable~\cite{Magdy:2017yje}
that was employed by STAR~\cite{Magdy:2020csm}.
We also investigate small system collisions simulated by AMPT. 
We then perform an ESE analysis using a toy model simulation in order to have sufficient statistics. 
We further examine the $R_{\Psi_{2}}$ variable with the toy model, 
investigating its effectiveness to identify the input CME signal and its vulnerability to  physics backgrounds, 
in an attempt to decipher the $R_{\Psi_{m}}$ variable.
We discuss our findings in the context of the STAR results~\cite{Magdy:2020csm}.

The rest of the article is organized as follows.
In Sec.~\ref{Methods}, the definitions of $R_{\Psi_{m}}$ and $\Delta\gamma$ are provided.
In Sec.~\ref{Results}, AMPT simulation results on $R_{\Psi_{m}}$ are presented in Au+Au, p+Au, and d+Au collisions.
In Sec.~\ref{SecEse}, an ESE study is conducted using toy model simulations.
In Sec.~\ref{ToyModelResults}, the toy model is used to study 
the elliptic flow ($v_{2}$) background and the CME signal ($a_{1}$) dependences for both  
$R_{\Psi_{2}}$ and $\Delta\gamma$ in Au+Au and isobar collisions.
In Sec.~\ref{Summary}, a summary is given.
Appendix~\ref{EPres} gives an analytical derivation for the event-plane resolution correction and discusses further complications.
In Appendix~\ref{CalcXi}, we extend our analytical analysis in Ref.~\cite{Feng:2018so} to the modified $R_{\Psi_{2}}$ variable for the pure background case, and derive an analytical form for the CME signal dependence of the $R_{\Psi_{2}}$ variable.
In Appendix~\ref{CalcDg}, we also provide an analytical form for the signal and background dependence of $\Delta\gamma$.


\section{Methodology} \label{Methods}

\subsection{The $R_{\Psi_{m}}$ correlator} \label{RmDefinition}

Phenomenologically, the azimuthal distribution of the primordial particles in each event can be expressed
into Fourier expansion
\begin{equation} \label{PrimoDist}
	\frac{\der N^{\pm}}{\der \phi} 
	\propto 1 \pm 2 a_{1} \sin(\phi - \Psi_{\text{RP}}) + 2 v_{2} \cos2(\phi-\Psi_{\text{RP}}) + \ldots
	,
\end{equation}
where $\Psi_{\text{RP}}$ denotes the RP azimuthal angle.
The $N^{\pm}$ is the number of particles with charge sign indicated by its superscript.
The coefficient $\pm a_{1}$ is the charge-dependent CME signal,
and $v_{2}$ is the elliptic flow coefficient.
In real data analysis, the RP is often surrogated by the second-order event plane (EP). The azimuthal angle of the EP of the order $m$ 
is calculated by
\begin{equation}
	\Psi_{m} = \frac{1}{m} \arctan\left(\frac{\sum_{i} w_{i} \sin(m\phi_{i})}{\sum_{i} w_{i} \cos(m\phi_{i})} \right)
	,
\end{equation}
where $\phi_{i}$ and $w_{i}$ are the azimuthal angle and weight of particle $i$.

To avoid auto-correlations, the particles of interests (POI, whose azimuth is $\phi$) to measure the CME (or the $a_1$ parameter)
must be excluded from the particles used to reconstruct the EP.
To realize that, the subevent method is used to define the $R_{\Psi_{m}}$ correlator.
Each event is divided into two subevents with a pseudorapidity gap 
-- one subevent (referred to as ``east'' subevent) with $-1.0<\eta<-0.1$ 
and the other (referred to as ``west'' subevent) with $0.1<\eta<1.0$.
We take the west subevent as an example to calculate the charge separation
perpendicular to the east-subevent EP ($\Delta S^{W}$) 
and parallel to it ($\Delta S^{\perp,W}$), according to the real charge sign. Namely,
\begin{equation} \label{SubDS}
\begin{split}
	\Delta S_{m}^{W} =& 
	\frac{1}{n_{W}^{+}} 
	\sum^{n_{W}^{+}}_{i \in W} \sin\left(\frac{m}{2}(\phi_{i}^{+} - \Psi_{m}^{E} ) \right) \\
	&- \frac{1}{n_{W}^{-}}
	\sum^{n_{W}^{-}}_{i \in W} \sin\left(\frac{m}{2}(\phi_{i}^{-} - \Psi_{m}^{E} ) \right) , \\
	\Delta S_{m}^{\perp,W} =& 
	\frac{1}{n_{W}^{+}} 
	\sum^{n_{W}^{+}}_{i \in W} \sin\left(\frac{m}{2}(\phi_{i}^{+} - \Psi_{m}^{E} ) + \frac{\pi}{2} \right) \\
	&- \frac{1}{n_{W}^{-}}
	\sum^{n_{W}^{-}}_{i \in W} \sin\left(\frac{m}{2}(\phi_{i}^{-} - \Psi_{m}^{E} ) + \frac{\pi}{2} \right)
	.
\end{split}
\end{equation}
To combine the two subevents, we take the average
\begin{equation} \label{AveSubDS}
\begin{split}
	\Delta S_{m} =& (\Delta S_{m}^{W} + \Delta S_m^{E})/2, \\
	\Delta S_{m}^{\perp} =& (\Delta S_{m}^{\perp,W} + \Delta S_m^{\perp,E})/2
	.
\end{split}
\end{equation}

The widths of the distributions in $\Delta S_m$ characterize the magnitude of charge separation with respect to the plane with which the $\Delta S_m$ is defined. The widths depend on the multiplicity of particles used to compute the $\Delta S_m$. To normalize out the multiplicity dependence, 
reference variables $\Delta S_{m, \text{sh}}$ and $\Delta S_{m, \text{sh}}^{\perp}$ are constructed by randomly shuffling the particle charge signs (according to relative abundances of positive and negative particles).
Denoting $\sigma_{m, \text{sh}}$ and $\sigma_{m, \text{sh}}^{\perp}$ for the RMS widths of the shuffled distributions, the $\Delta S_{m}$ variables are scaled as follows~\cite{Magdy:2020csm},
\begin{equation}
\Delta {S'}_{m} = \Delta S_{m} / \sigma_{m, \text{sh}}, \quad
	\Delta {S'}_{m}^{\perp} = \Delta S_{m}^{\perp} / \sigma_{m, \text{sh}}^{\perp}.
\end{equation}
Because of finite multiplicity fluctuations, the reconstructed EP is smeared from the RP, broadening the $\Delta S_m$ distributions. A multiplicative factor is applied to correct for the effect of the imperfect EP reconstruction,
\begin{equation} \label{ScaledDS}
	\Delta {S''}_{m} = \Delta {S'}_{m} \delta_{r_{m}}, \quad
	\Delta {S''}_{m}^{\perp} = \Delta {S'}_{m}^{\perp} \delta_{r_{m}}.
\end{equation}
The correction factor is given by 
\begin{equation}\label{eq:rm1}
\delta_{r_{m}} = \sqrt{r_{m}},
\end{equation}
where $r_m$ is the EP resolution of subevents,
\begin{equation} \label{SubEpRes}
\begin{split}
	r_{m} = \langle \cos m(\Psi_{m}^{E/W}-\Psi_{\text{RP}}) \rangle = \sqrt{\langle \cos m(\Psi_{m}^{W}-\Psi_{m}^{E}) \rangle}
	.
\end{split}
\end{equation}
The derivation of Eq.~\ref{eq:rm1} is given in Appendix~\ref{EPres}.\ref{EPresCalc}.

The normalized distributions of $\Delta {S''}_{m}$ are
\begin{equation} \label{CCorr}
\begin{split}
	C_{\Psi_{m}} = \frac{\text{event probability distribution in } \Delta {S''}_{m}} { \text{event probability distribution in } \Delta {S''}_{m,\text{sh}}} , \\
	C_{\Psi_{m}}^{\perp} = \frac{\text{event probability distribution in } \Delta {S''}_{m}^{\perp}} { \text{event probability distribution in } \Delta {S''}_{m,\text{sh}}^{\perp}}
	.
\end{split}
\end{equation}
The $R_{\Psi_{m}}$ observable is defined by the double ratio
\begin{equation} \label{RCorr}
	R_{\Psi_{m}} = \frac{ C_{\Psi_{m}} } { C_{\Psi_{m}}^{\perp} }
	.
\end{equation}
We characterize the shape of $R_{\Psi_{m}}$ by 
\begin{equation} \label{kal}
\begin{split}
	&\kal = -
	\frac{1}{\delta_{r_{m}}^{2}} \left(\frac{\sigma_{m,\text{sh}}^{2}}{\sigma_{m}^{2}}-\frac{{\sigma_{m,\text{sh}}^{\perp}}^{2}}{{\sigma_{m}^{\perp}}^{2}} \right)
	.
\end{split}
\end{equation}
The variable $\kal$ can also be obtained by fitting the $R_{\Psi_{m}}$ distributions to $C e^{\kal x^{2}/2}$.
The distribution of $R_{\Psi_{m}}$ is concave when $\kal>0$, convex when $\kal<0$, and flat when $\kal=0$.
The width of $R_{\Psi_{m}}$ is $\sigma = 1/\sqrt{|\kal|}$,
so we will refer to $\kal$ as the squared inverse width of $R_{\Psi_{m}}$.

As will be discussed in Section~\ref{V2Dependence}, the averaging of Eq.~\ref{AveSubDS} introduces auto-correlations and is thus not a good way to define $\Delta S_m$ and $\Delta S_m^{\perp}$. 
We propose not to average the two subevents but treat them separately.
See Section~\ref{V2Dependence} and Appendix~\ref{EPres} for more details. 
Nonetheless, for comparisons to the previous works, we study both cases where the subevents are averaged as well as treated independently, and use the same correction factor given by Eq.~\ref{eq:rm1}.

\subsection{The $\Delta\gamma$ observable} \label{GammaDFDefinition}

The two-particle azimuthal correlator $\Delta\gamma$ observable~\cite{Voloshin:2004vk} is widely used in CME studies at RHIC~\cite{Abelev:2009ac,Abelev:2009ad,Adamczyk:2014mzf,Adamczyk:2013hsi,Zhao:2017wck,Zhao:2017ckp}
and the LHC~\cite{Abelev:2012pa,Khachatryan:2016got,Sirunyan:2017quh,Acharya:2017fau, Acharya:2020rlz}.
For completeness, we give a brief description of the $\Delta\gamma$ observable. 
To keep consistency with the $R_{\Psi_{m}}$, 
we define $\Delta\gamma$ also by subevents,
\begin{equation} \label{subgamma}
\begin{split}
	\gamma_{\text{OS}} =& \langle \cos(\phi_{a \in E/W}^{\pm} + \phi_{b \in E/W}^{\mp} - 2 \Psi_{2}^{W/E}) \rangle 
	\left/ r_{2} \right. , \\
	\gamma_{\text{SS}} =& \langle \cos(\phi_{a \in E/W}^{\pm} + \phi_{b \in E/W}^{\pm} - 2 \Psi_{2}^{W/E}) \rangle 
	\left/ r_{2} \right. , \\
	\Delta\gamma =& \gamma_{\text{OS}} - \gamma_{\text{SS}},
\end{split}
\end{equation}
where $a$ and $b$ are two particles in the same subevent and
$r_{2}$ is the second-order EP resolution of the subevents, as in Eq.~\ref{SubEpRes}.
In order to compare with $R_{\Psi_{2}}$, the same POI cuts and EP particle cuts as in $R_{\Psi_{2}}$ are used in $\Delta\gamma$.

For the CME signal parameterized by the $a_1$ parameter in Eq.~\ref{PrimoDist}, 
the $\Delta\gamma$ correlator can be obtained as
\begin{equation} \label{GammaDFa1}
    \Delta\gamma=2a_1^2.
\end{equation}
It is well known that $\Delta\gamma$ is strongly contaminated by physics backgrounds caused by two-particle correlations and the anisotropy of those correlated pairs~\cite{Voloshin:2004vk, Wang:2009kd, Adamczyk:2013kcb,  Bzdak:2009fc,Schlichting:2010qia}. For instance, resonance decays present a major background:
\begin{equation} \label{GammaDFv2}
    \Delta\gamma=\frac{N_{\rm reso}}{N_{\rm pair}}\langle\cos(\phi_a+\phi_b-2\phi_{\rm reso})\rangle v_{2,\rm{reso}}.
\end{equation}
Since the number of resonances $N_{\rm reso}\propto N$, the number of pairs $N_{\rm pair}\propto N^2$, and the resonance elliptic flow $v_{2,{\rm reso}}\propto v_2$, the background contamination in $\Delta\gamma$ is generally proportional to the final-state particle $v_2$ and inversely proportional to the multiplicity ($N$).


\section{AMPT results} \label{Results}

The AMPT model~\cite{Lin:2004en} is widely used 
to simulate relativistic heavy ion collisions, without CME signal.
In this study, we use the AMPT version v2.25t4cu2 where charge conservation is ensured.
We set the model parameter NTMAX=150 which means that the hadronic cascade is turned on.
For particles used for EP reconstruction, 
a cut is applied to their transverse momentum $0.2 \text{ GeV/c} < p_{T} < 2.0 \text{ GeV/c}$,
while for POI, a tighter $p_{T}$ cut is applied $0.35 \text{ GeV/c} < p_{T} <2.0 \text{ GeV/c}$, as in the STAR analysis~\cite{Magdy:2020csm}.
All particles used in our analysis are required to be inside the $\eta$ range $-1<\eta<1$.

\begin{figure}
	\includegraphics[width=1.0\linewidth]{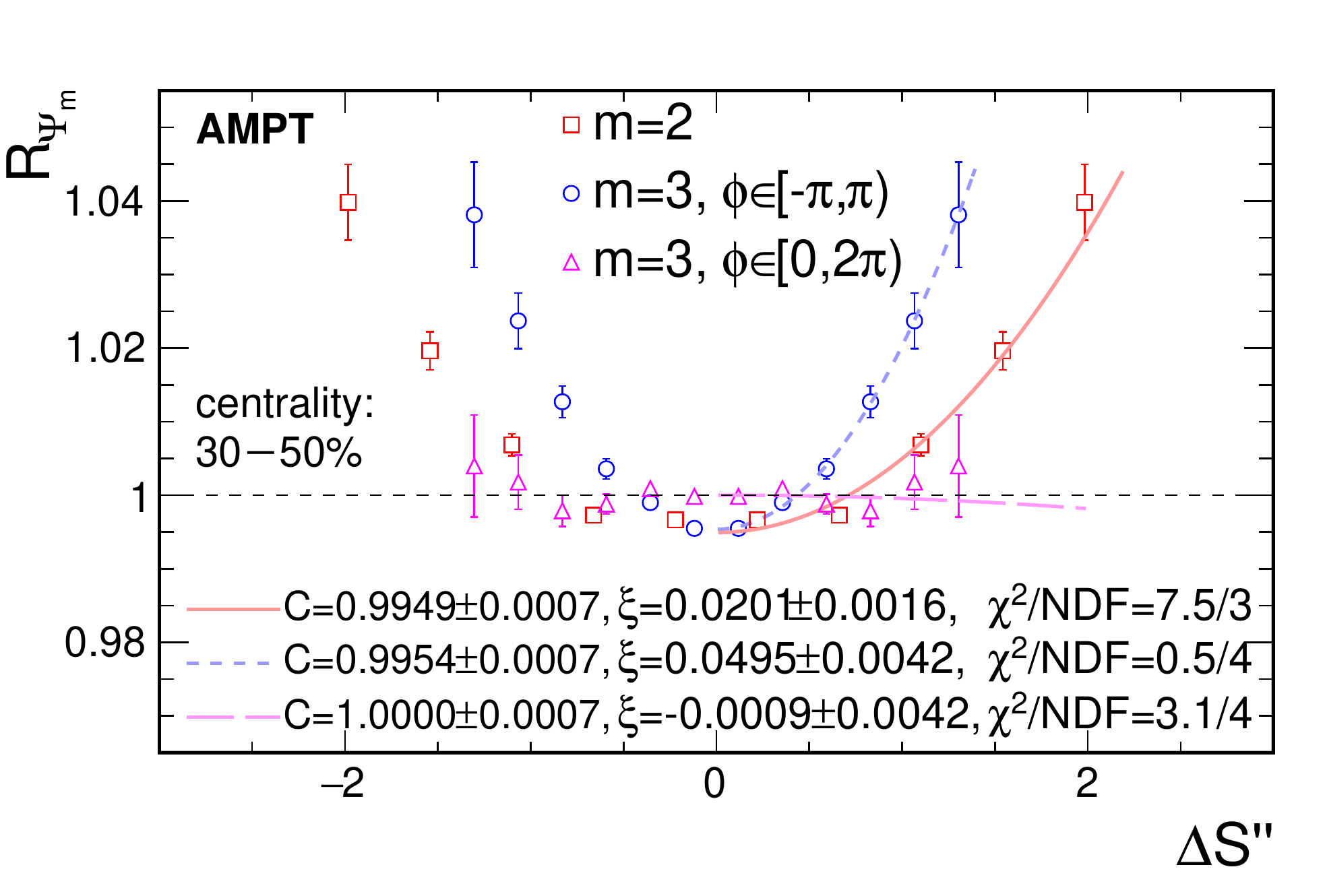}
	\caption{The $R_{\Psi_{2}}$ and $R_{\Psi_{3}}$ distributions in centrality 30--50\% 
	Au+Au collisions at $\sqrt{s_{NN}}=200 \text{ GeV}$ simulated by AMPT.
	Total 25.8 million 30--50\% centrality events are generated and analyzed.
	The POI are required to have $0.35\text{ GeV/c} < p_{T} < 2.0 \text{ GeV/c}$ and $0.1< | \eta | <1.0$,
	whereas particles used for EP reconstruction are required to have 
	$0.2 \text{ GeV/c} < p_{T} < 2.0 \text{ GeV/c}$ and $0.1< | \eta | <1.0$.
	The $R_{\Psi_{m}}$ distributions are symmetrized.
	With the azimuth range $\phi\in[-\pi,\pi)$, the $R_{\Psi_{2}}$ (red square) and $R_{\Psi_{3}}$ (blue circle) curves are both concave and apparently different.
	With the range $\phi \in [0,2\pi)$, $R_{\Psi_{3}}$ (magenta triangle) is relatively flat and $R_{\Psi_{2}}$ is unchanged.
	The curves are fits to function $f(x) = C e^{ \kal x^{2}/2}$.}
	\label{AmptR2R3}
\end{figure}

\begin{figure}
	\includegraphics[width=1.0\linewidth]{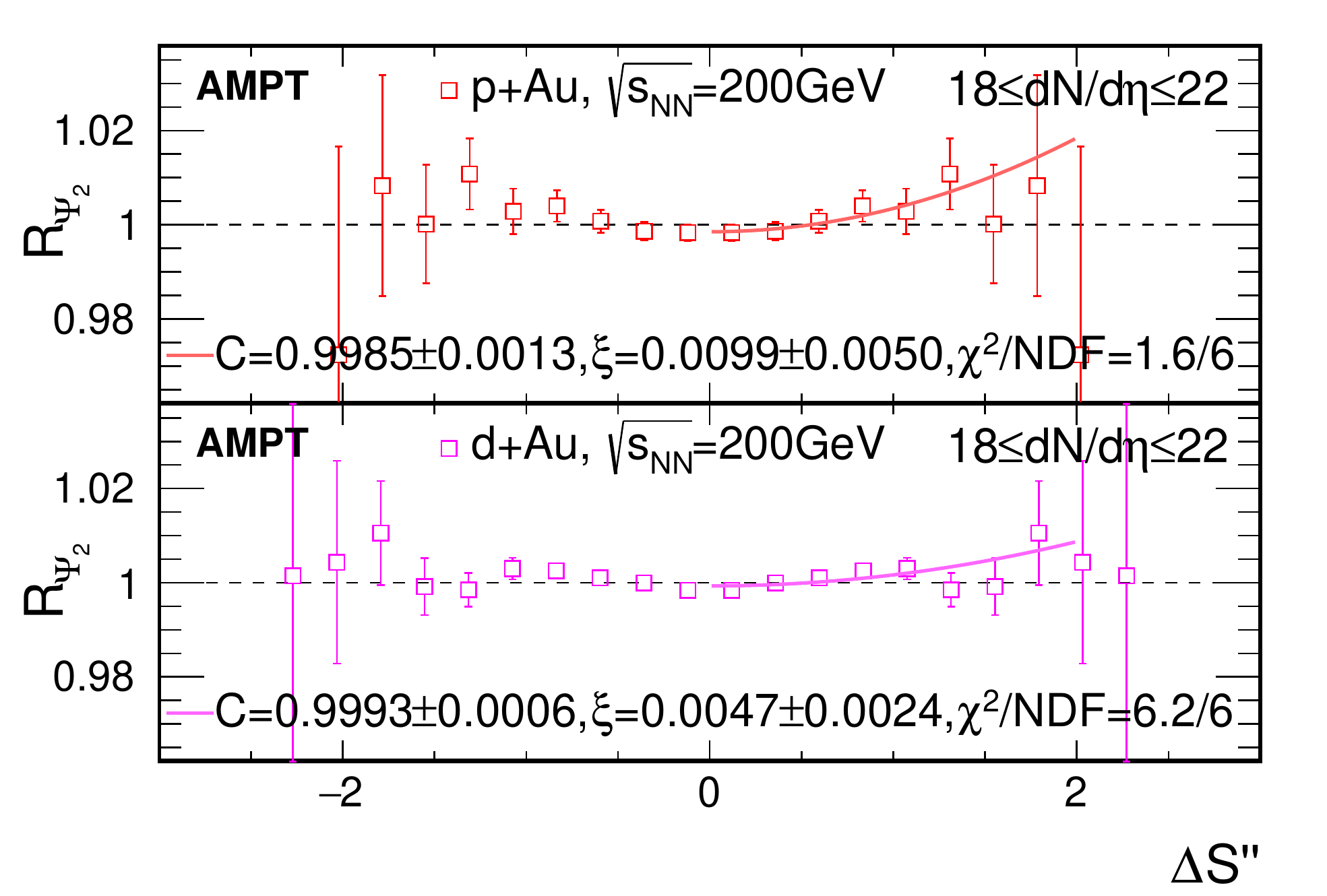}
	\caption{$R_{\Psi_{2}}$ distributions in $18 \le \refm \le 22$ from AMPT
	simulations for p+Au collisions (upper pad) and d+Au collisons (lower pad) at $\sqrt{s_{NN}}=200 \text{ GeV}$.
	Each dataset has 300 million MB events; $7.3$ and $33.2$ million analyzed p+Au and d+Au events in the $\refm$ range, respectively.
	The POI are required to have $0.35\text{ GeV/c} < p_{T} < 2.0 \text{ GeV/c}$ and come from the p/d-going range $0.1<\eta<1.0$,
	whereas the particles used for EP reconstruction are required to have $0.2 \text{ GeV/c}<p_{T}<2.0 \text{ GeV/c}$ 
	and come from the Au-going range $-1.0<\eta<-0.1$.
	The $R_{\Psi_{2}}$ distributions are symmetrized.
	The $R_{\Psi_{2}}$ from both p+Au and d+Au are slightly concave.
	The curves are fits to function $f(x) = C e^{ \kal x^{2}/2}$.}
	\label{AmptSmallSys}
\end{figure}

\subsection{Au+Au collisions} \label{AmptAuAu200GeV}

For Au+Au collisions at $\sqrt{s_{NN}} = 200 \text{ GeV}$,
the minimal bias (MB) AMPT events are generated first 
to define centrality by cutting on MB multiplicity distribution.
Then, a total of $25.8$ million 30--50\% centrality events are simulated and used for this analysis.
Figure~\ref{AmptR2R3} shows the $R_{\Psi_{2}}$ (red square) and $R_{\Psi_{3}}$ (blue circle) distributions.
The distributions are concave and different from each other in width.
This is in stark contrast to the STAR results in Ref.~\cite{Magdy:2020csm},
where convex, nearly-identical  $R_{\Psi_{2}}$ and $R_{\Psi_{3}}$ curves were obtained.
The ``identical'' $R_{\Psi_{2}}$ and $R_{\Psi_{3}}$ curves from AMPT 
(where only backgrounds are present, with no CME singal) 
was critical for the claim in Ref.~\cite{Magdy:2020csm} that the Au+Au data, 
where different $R_{\Psi_{2}}$ and $R_{\Psi_{3}}$ distributions are observed, 
are consistent with CME and inconsistent with known backgrounds. 
Since the $v_{2}$ and $v_{3}$ physics mechanisms are the same in the hydrodynamic picture, 
it may be satisfactory to find identical $R_{\Psi_{2}}$ and $R_{\Psi_{3}}$ curves. 
However, our AMPT results in Fig.~\ref{AmptR2R3} demonstrate that
the $R_{\Psi_{2}}$ and $R_{\Psi_{3}}$ are not necessarily the same
when only pure background is present.
We speculate that the difference roots in the $R_{\Psi_{m}}$ definitions: 
the ``harmonic'' multiplier $m/2$ in front of the azimuthal angle $\phi-\Psi_{m}$ (see Eq.~\ref{SubDS}) 
renders actually two distinctly different variables of $R_{\Psi_{2}}$ and $R_{\Psi_{3}}$. 

Moreover, as pointed out in Ref.~\cite{Feng:2018so}, 
the $R_{\Psi_{3}}$ variable is ill-defined because it breaks the natural azimuthal periodicity of $2\pi$. 
The $R_{\Psi_{3}}$ in blue circles in Fig.~\ref{AmptR2R3} uses the azimuthal range of $\phi \in [-\pi,\pi)$. 
If it is switched to $\phi \in [0,2\pi)$ by adding $2\pi$ to those in the range $[-\pi,0)$, with no change in physics, minus signs appear to the corresponding terms in Eq.~\ref{SubDS}, and the $R_{\Psi_{3}}$ distribution changes completely to the magenta triangles in Fig.~\ref{AmptR2R3}.
The $R_{\Psi_{2}}$ is of course unchanged by the choice of the $\phi$ range.
Since $R_{\Psi_{3}}$ is ill-defined~\cite{Feng:2018so},
we will only focus on $R_{\Psi_{2}}$ in the rest of this paper.

We note that a recent publication~\cite{MNML} appeared 
with similar AMPT results as those in the STAR work~\cite{Magdy:2020csm}. 
An examination of the statistical errors suggests~\cite{Feng:2020com} 
that those AMPT results in Ref.~\cite{MNML} are highly improbable to be real,
calling into question the validity of those AMPT results.
Moreover, concave $R_{\Psi_{m}}$ distributions were observed by several other model studies
for Au+Au collisions at $\sqrt{s_{NN}} = 200 \text{ GeV}$.
Those include hydrodynamic simulations~\cite{Bozek:2018aad} 
and toy model studies~\cite{Feng:2018so}.

\subsection{p+Au and d+Au collisions}

For the small systems p+Au and d+Au collisions at $\sqrt{s_{NN}} = 200 \text{ GeV}$, total 300 million MB AMPT events each are simulated.
Since the centrality is not well-defined in those small systems,
we cut on the reference multiplicity $18 \le \refm \le 22$ (the number of charged particles in the range $-0.5 < \eta < 0.5$), a range similar to the STAR data analysis~\cite{Magdy:2020csm}. These correspond to $7.3$ and $33.2$ million analyzed events for p+Au and d+Au collisions, respectively.
As same as in Ref.~\cite{Magdy:2020csm},
the event plane is reconstructed from the particles in the Au-going direction in the range of $-1.0<\eta<-0.1$,
and the POI's are from the p/d-going direction in the range of $0.1<\eta<1.0$. The $\eta$ gap between the EP particles and the POI's suppresses short-range correlations.

Figure~\ref{AmptSmallSys} shows the $R_{\Psi_{2}}$ distributions in the small system collisions by AMPT. 
The distributions are slightly concave, and appear 
qualitatively different from the STAR data~\cite{Magdy:2020csm},
where the $R_{\Psi_{2}}$ curve in p+Au collisions is flat and that in d+Au collisions is flat or even convex. 
Since the CME signal is either absent or uncorrelated with the reconstructed EP in those small systems, 
the flat $R_{\Psi_{2}}$ curves were important for the conclusion in Ref.~\cite{Magdy:2020csm}
that the $R_{\Psi_{2}}$ is sensitive to CME and relatively insensitive to backgrounds 
which do have strong effects on the $\Delta\gamma$ observable~\cite{Zhao:2017wck,Zhao:2017ckp}. 
Our AMPT results in p+Au and d+Au collisions suggest that this may not be the case.


\begin{figure}
	\includegraphics[width=0.8\linewidth]{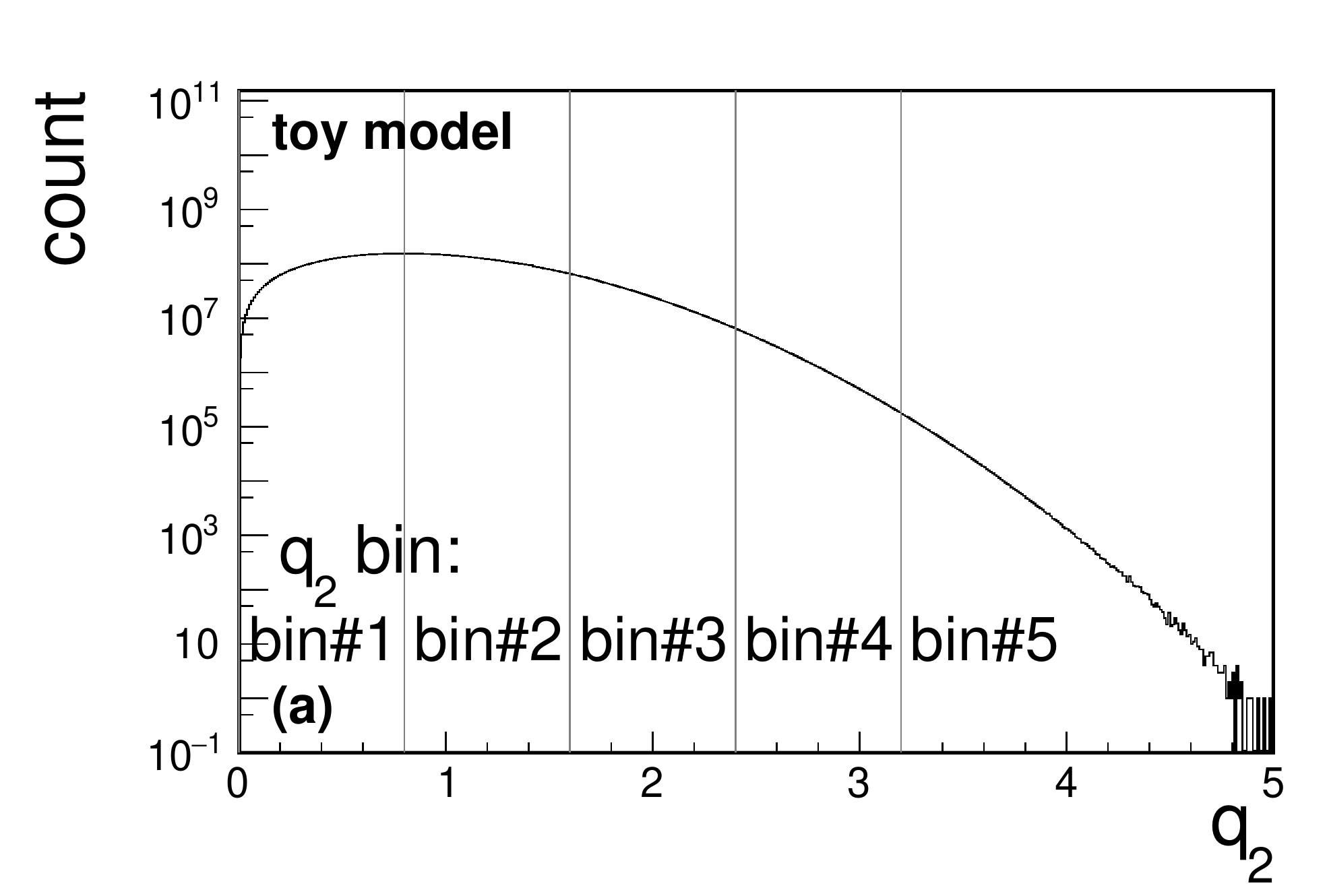}
	\includegraphics[width=0.8\linewidth]{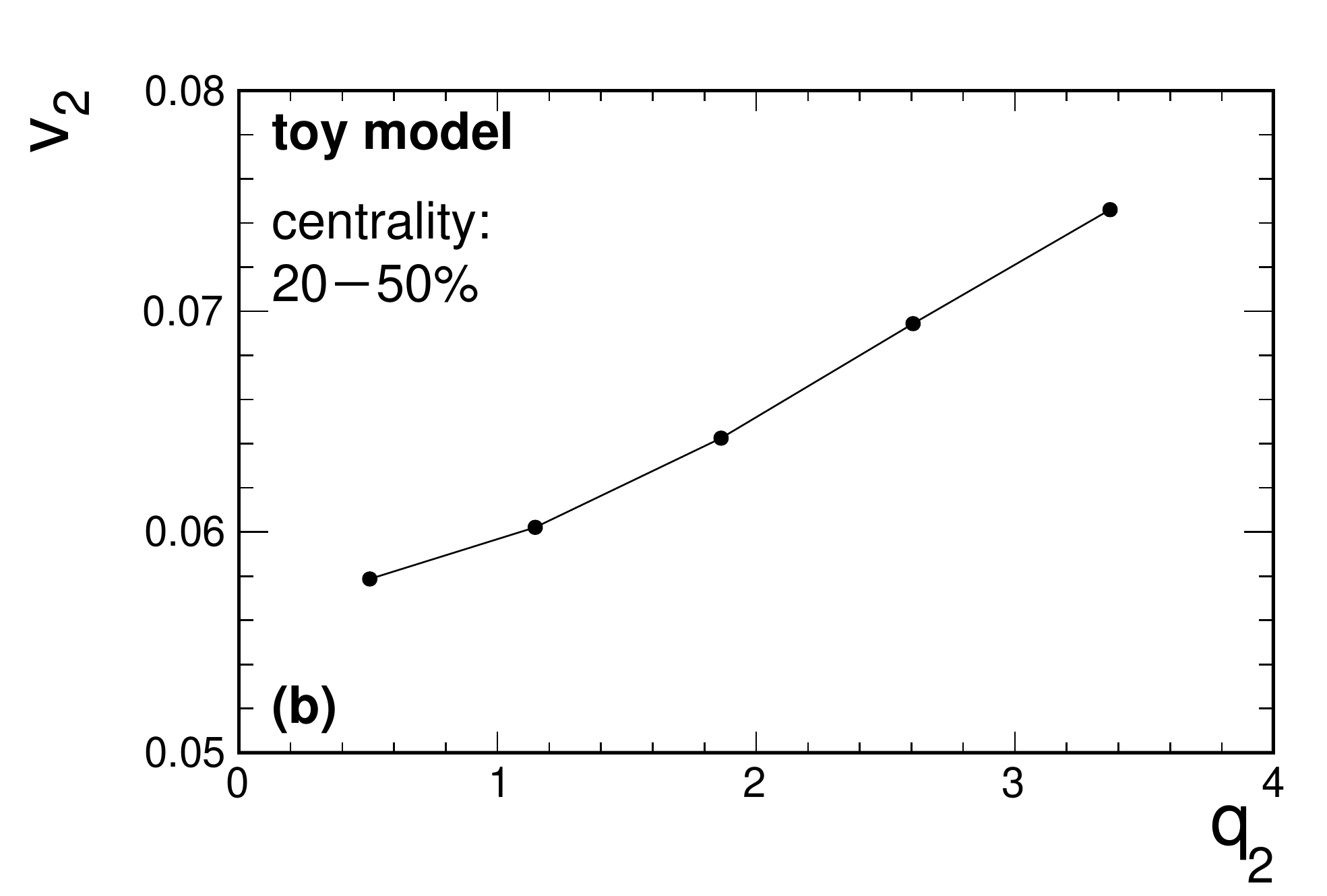}
	\caption{(a) The $q_{2}$ distribution for 20--50\% Au+Au collisions 
	simulated by the toy model using STAR data as input parameters.
	This is used in ESE analysis where 
	$q_{2}$ bins are divided by $q_{2}$ values of equal spacing (except the last bin).
	(b) The $\vvese$ vs.~$q_{2}$ of each ESE $q_{2}$ bin.
	Total 10.9 billion events are simulated for centrality 20--50\%.
	The $q_{2}$ is calculated from particles 
	in $| \eta |<0.3$ with $0.2 \text{ GeV/c} < p_{T} < 2.0 \text{ GeV/c}$,
	whereas $\vvese$ is calculated from particles 
	in $| \eta |>0.3$ with $0.2 \text{ GeV/c} < p_{T} < 2.0 \text{ GeV/c}$.
	}
	\label{Q2DistComp}
\end{figure}

\section{ESE study in a toy model} \label{SecEse}

\begin{figure}
	\includegraphics[width=0.8\linewidth]{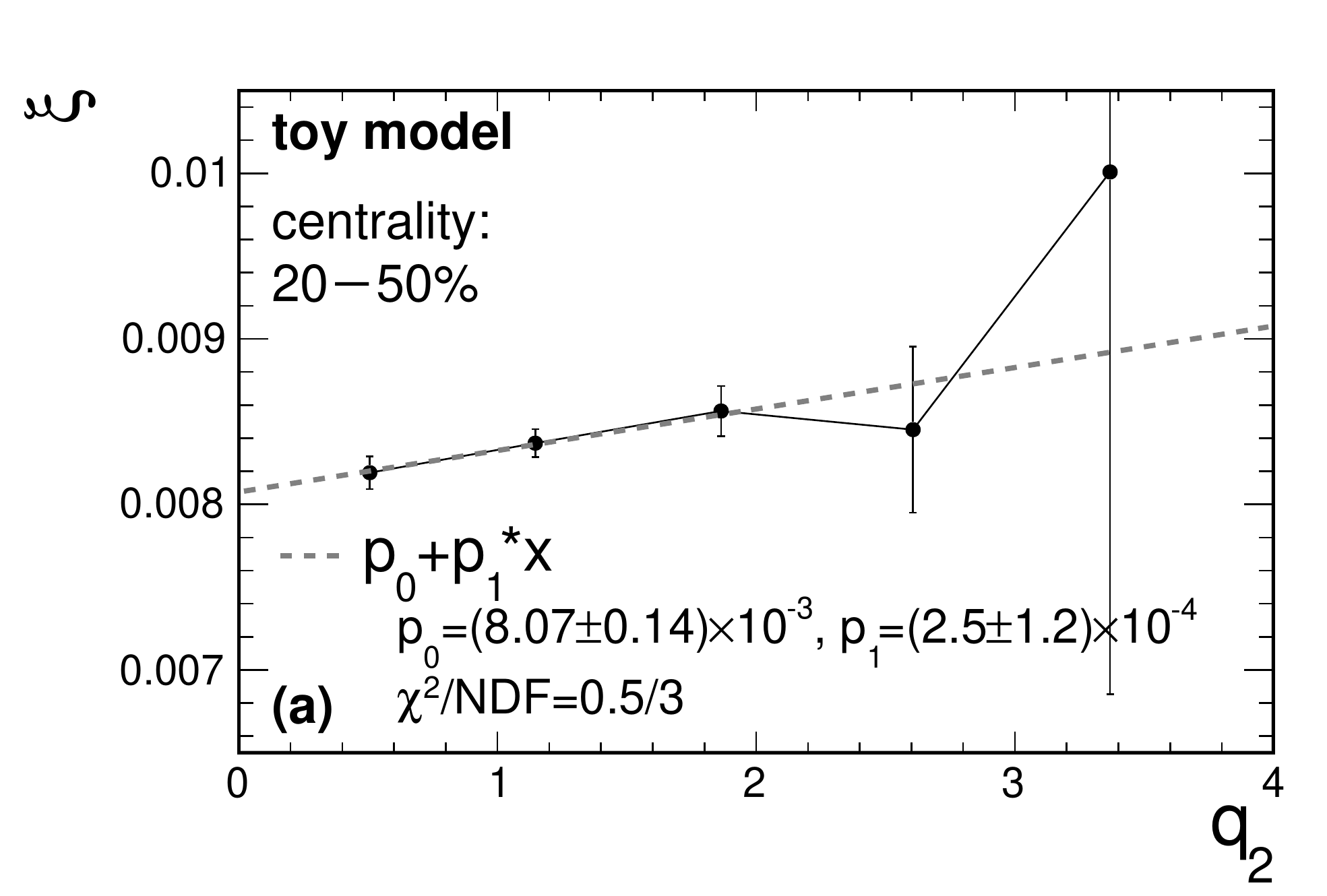}
	\includegraphics[width=0.8\linewidth]{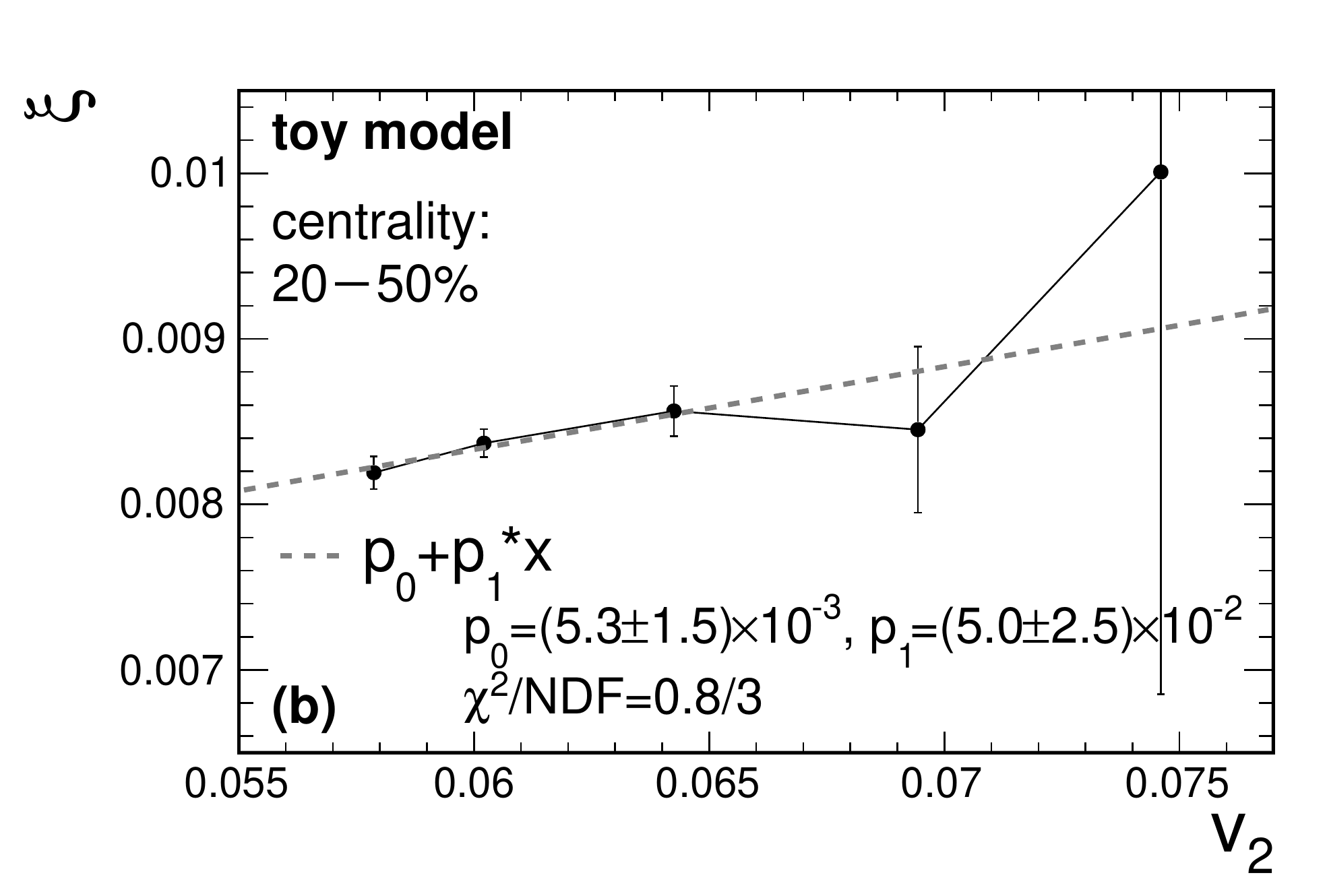}
	\caption{(a) The $\kal$ vs.~$q_{2}$ of each ESE $q_{2}$ bin
	for centrality 20--50\% Au+Au collisions 
	simulated by the toy model using STAR data as input parameters.
	Total 10.9 billion events are simulated for centrality 20--50\%.
	The POI are required to have $0.35\text{ GeV/c} < p_{T} < 2.0 \text{ GeV/c}$ and $0.3< \pm\eta <1.0$,
	whereas particles for EP reconstruction are required to have 
	$0.2 \text{ GeV/c} < p_{T} < 2.0 \text{ GeV/c}$ and $0.3< \mp\eta <1.0$.
	(b) The mapping of $\kal$ as a function of $\vvese$ in each ESE $q_{2}$ bin.
	}
	\label{SumXEse}
\end{figure}

STAR performed an ESE analysis of their Au+Au data~\cite{Magdy:2020csm}. 
Each event is divided into three subevents: 
east ($-1<\eta<-0.3$), middle ($-0.3<\eta<0.3$) and west ($0.3<\eta<1.0$) subevents.
The middle subevent is used to calculate the $q_{2}$ quantity,
\begin{equation} \label{q2}
	q_{2} = \sqrt{\frac{  \left( \sum_{i}^{M} \cos2\phi_{i} \right)^{2} + \left( \sum_{i}^{M} \sin2\phi_{i} \right)^{2} }{M}}
	,
\end{equation}
where 
$M$ is the number of particles in the middle subevent.
This quantity is related to the elliptical shape of the corresponding subevent in momentum space.
The events are then divided according to the $q_2$ value, and are analyzed separately in each $q_2$ class.
In each event, the east and west subevents are used to calculate the elliptic flow $v_{2}$
and the $R_{\Psi_{2}}$ correlator by the subevent method (Eqs.~\ref{SubDS}--\ref{RCorr}).
It was found that the $v_2$ increases with increasing $q_2$ but the $R_{\Psi_{2}}$ width is independent of $q_2$ within uncertainties.
This would imply that the width of $R_{\Psi_{2}}$ 
is independent from the event-by-event $\vvese$ in each $q_{2}$ class. 
This renders support to the claim in Ref.~\cite{Magdy:2020csm} that the $R_{\Psi_{2}}$ is relatively insensitive to the flow background.

\begin{figure*}
	\includegraphics[width=0.4\linewidth]{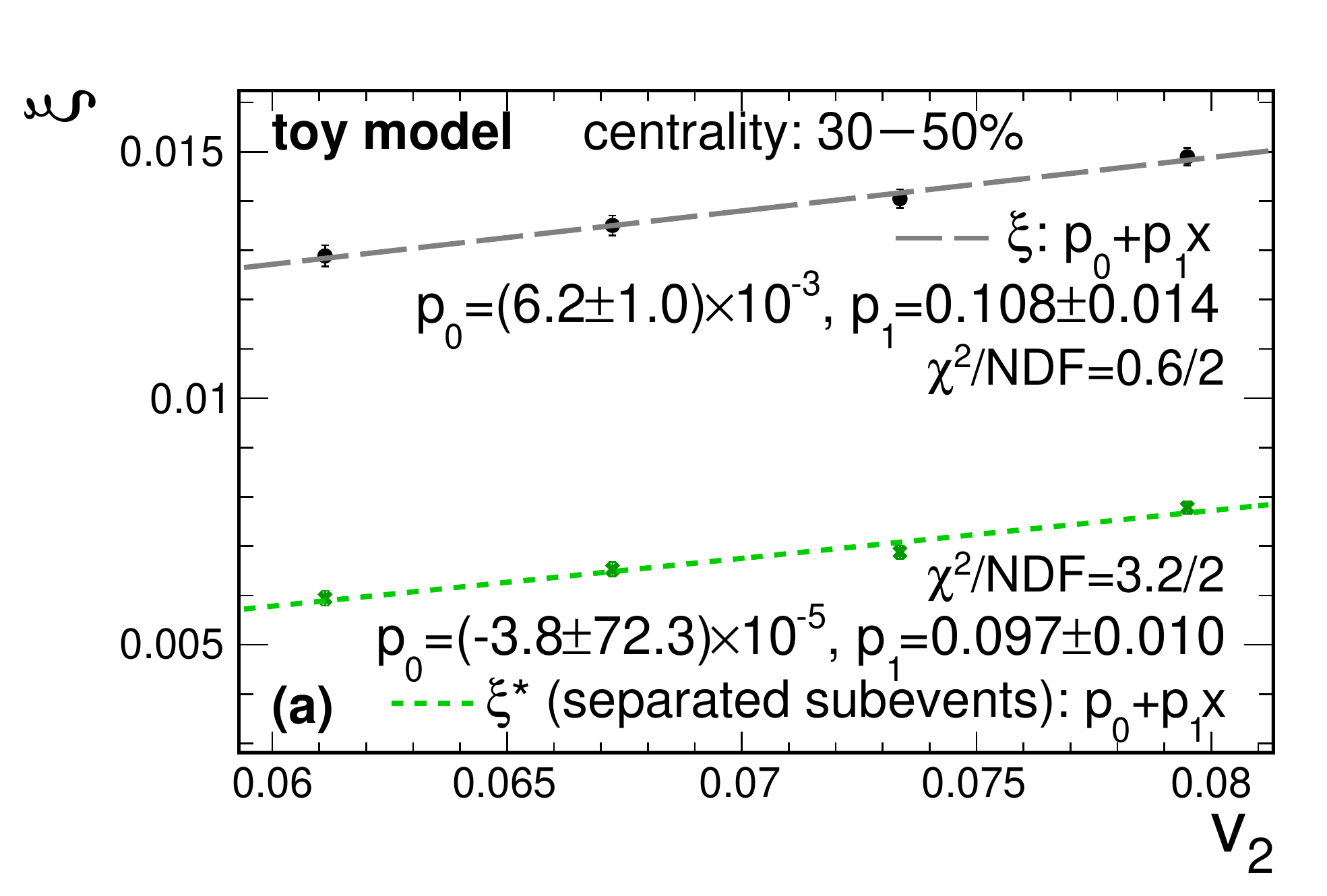}
	\includegraphics[width=0.4\linewidth]{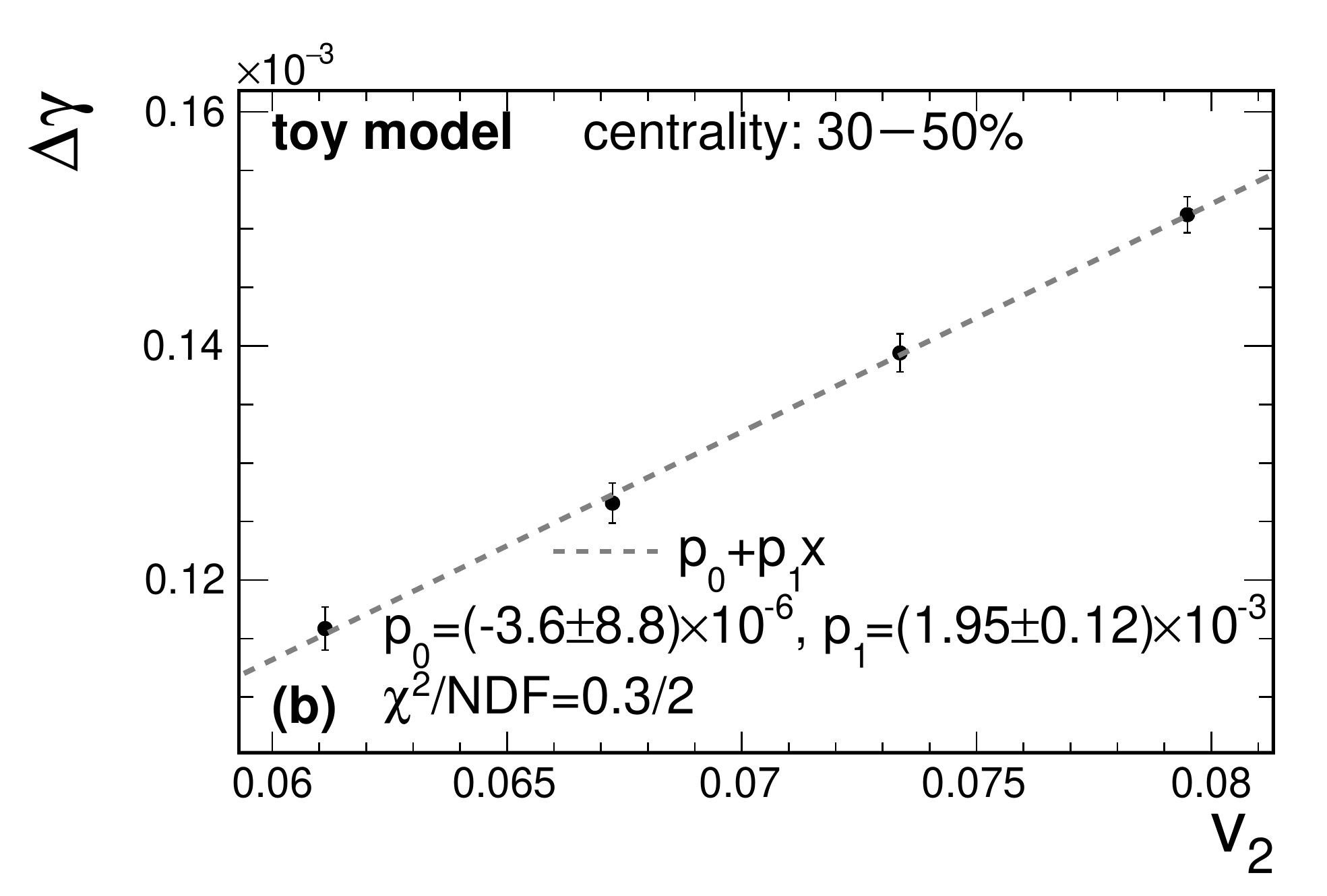}
	\caption{
	The $\kal$ and $\skal$ (a) and $\Delta\gamma$ (b) as functions of the input $v_{2}$ from 
	Au+Au collisions at $\sqrt{s_{NN}}=200 \text{ GeV}$ with no CME signal.
	The leftmost data point uses the default $v_{2}$ distributions; 
	other points use $v_{2}$ distributions that are scaled up accordingly.
	Each data point has total 2 billion toy model events in the 0--80\% centrality range
	(or $0.73$ billion events in the 30--50\% centrality range).
	The POI are required to have $0.35\text{ GeV/c} < p_{T} < 2.0 \text{ GeV/c}$ and $0.1< \pm \eta <1.0$,
	whereas particles for EP reconstruction are required to have 
	$0.2 \text{ GeV/c} < p_{T} < 2.0 \text{ GeV/c}$ and $0.1< \mp \eta <1.0$.
	For each dataset,
	the $\kal$, $\skal$, $\Delta\gamma$,
	and $v_{2}$ are calculated from centrality range 30--50\% 
	(average multiplicity $\refm \approx 140$, 
	average subevent POI multiplicity $\mult \approx 64$).
	The dashed lines are two-parameter linear fits.
	}
	\label{ToyAveV2InvWth}
\end{figure*}

\begin{figure*}
	\includegraphics[width=0.325\linewidth]{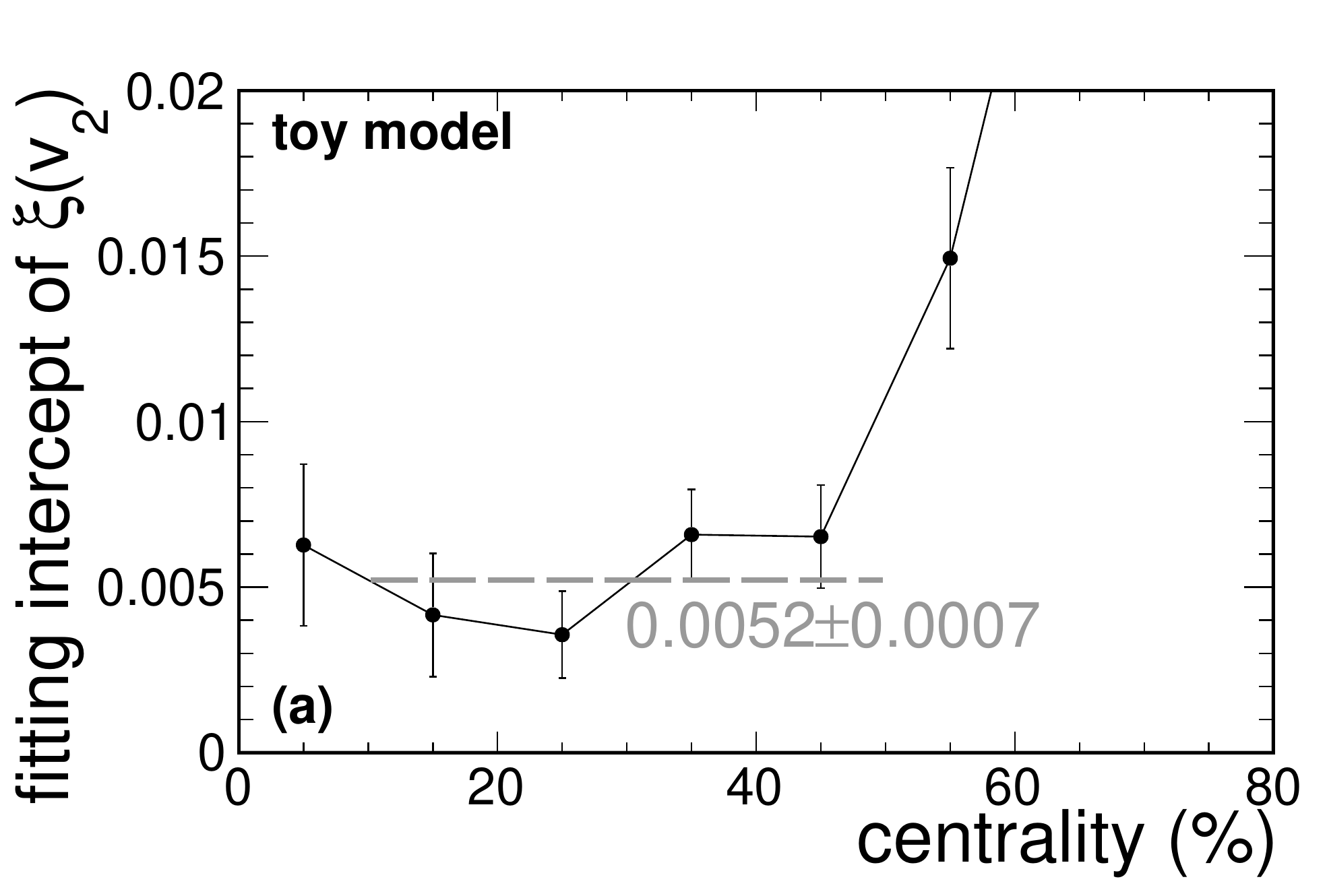}
	\includegraphics[width=0.325\linewidth]{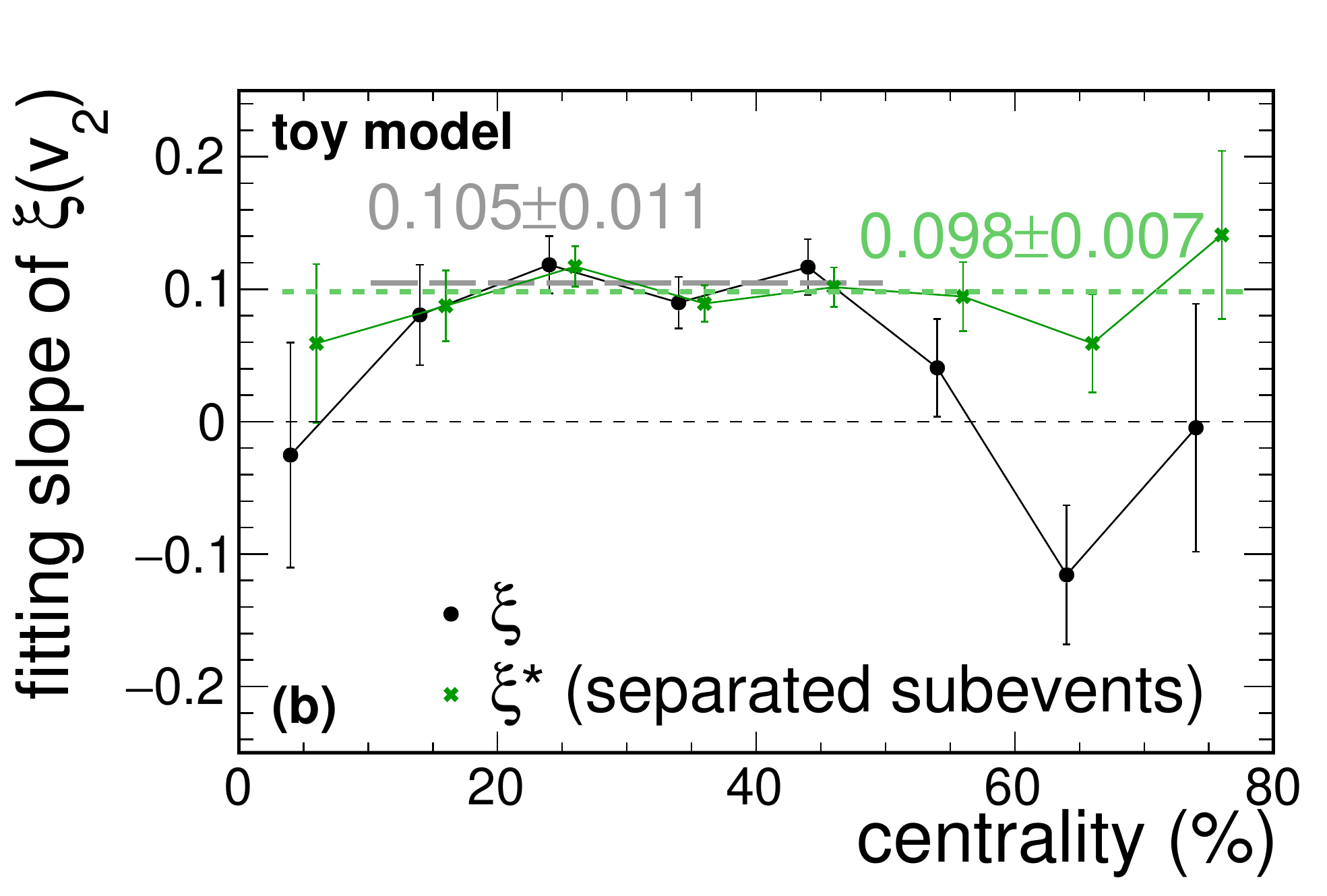}
	\includegraphics[width=0.325\linewidth]{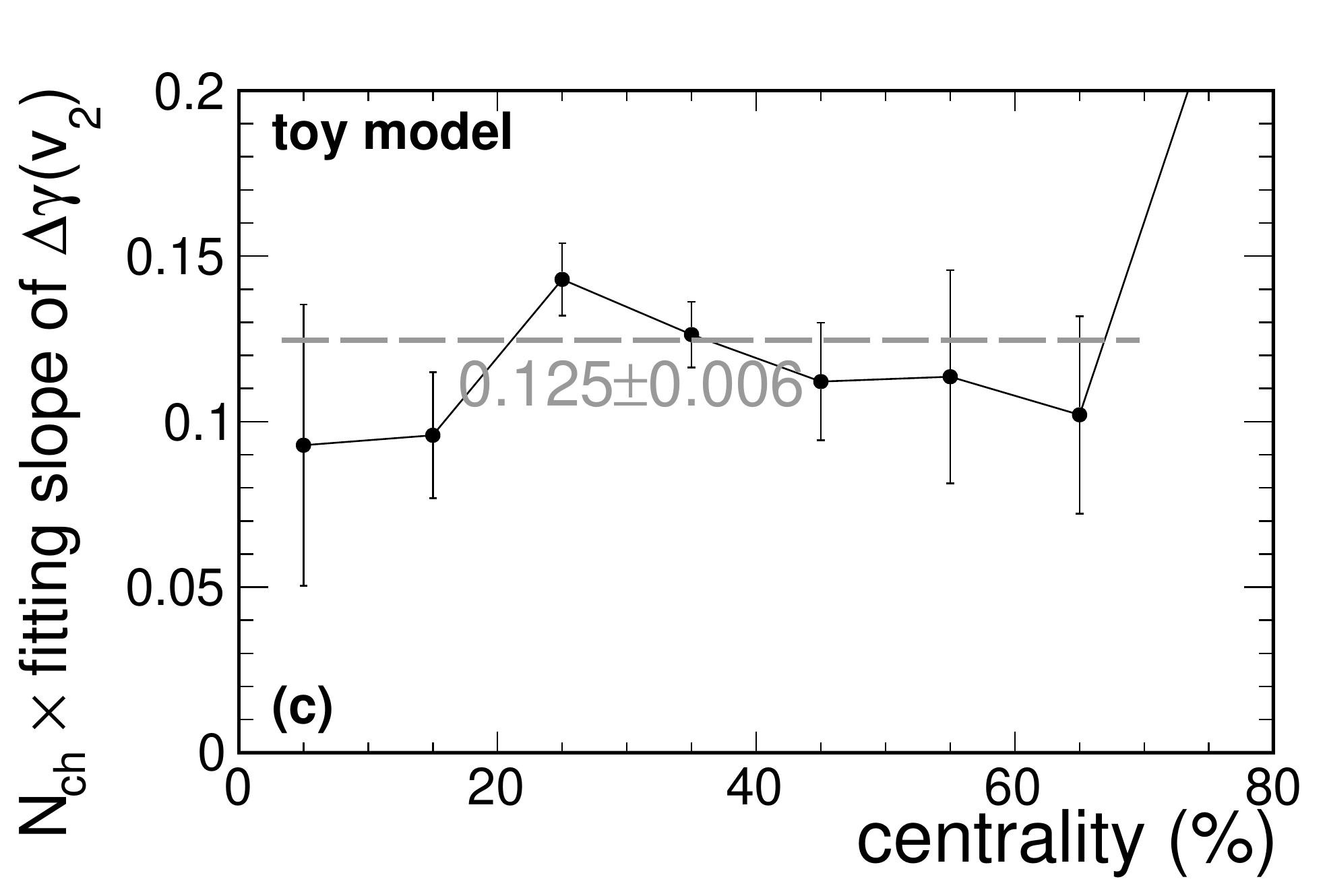}
	\caption{
	The fit parameters from linear fits to $\kal(v_{2})$, $\skal(v_{2})$, and $\Delta\gamma(v_{2})$ 
	are plotted as functions of centrality.
	The $\kal$, $\skal$, $\Delta\gamma$ are similar to those in Fig.~\ref{ToyAveV2InvWth}, 
	but are calculated for each narrow centrality bin of 10\% size.
	(a) The $\kal(v_{2})$ intercept, which is roughly constant in the centrality range 10--50\%.
	The $\skal(v_{2})$ and $\Delta\gamma(v_{2})$ intercepts are all consistent with zero. 
	(b) The slopes of $\kal(v_{2})$ and $\skal(v_{2})$,
	which seem to be independent of centrality.
	(c) The $\Delta\gamma(v_{2})$ slope multiplied by $\mult$. 
	The slope is inversely proportional to $\mult$.
	}
	\label{V2Cent}
\end{figure*}

However, our previous toy model study~\cite{Feng:2018so} shows that 
$R_{\Psi_{2}}$ has dependence on event-wise $v_{2}$.
This seems to contradict the claim from the ESE study in Ref.~\cite{Magdy:2020csm}.
To investigate this further, 
we carry out an ESE analysis using the toy model~\cite{Wang:2016iov, Feng:2018so}. 
The toy model is used instead of a physics model such as AMPT because the ESE analysis typically requires large statistics that is difficult to achieve by the latter.
The toy model includes primordial pions and $\rho$ meson decay daughters~\cite{Wang:2016iov, Feng:2018so}.
The inputs to the toy model are taken from real data of Au+Au collisions at $200$ GeV 
for each of the $10\%$-size centrality bins. 
These include the pion and $\rho$ meson $p_T$ distributions and $v_2(p_T)$~\cite{Adams:2003cc, Adler:2003qi, Adams:2003xp, Abelev:2008ab, Adams:2004bi, Adare:2010sp, Dong:2004ve, Adamczyk:2015lme, Agashe:2014kda, Abelev:2009gu, Wang:2016iov}.
The $p_{T}$ spectra measurements of the $\rho$ mesons are limited to 40--80\% centrality;
the $p_{T}$ spectra shapes are assumed to be centrality independent in our simulation. 
The $\vvr(p_{T})$ are parameterized according to the number of constituent quark (NCQ) scaling.
Fluctuations are added for $v_{2}$ by a Gaussian distribution with a relative width of $40\%$ from event to event.
The $\rho/\pi$ multiplicity ratio is approximately $\nr/\np=0.085$, assumed to be centrality independent, 
and their multiplicities are such 
that the total final multiplicity matches mid-rapidity data for each given centrality bin~\cite{Wang:2016iov}. 
Particles are generated with $|\eta|<1.5$ assuming multiplicity density is uniform in $\eta$.
For a given centrality bin of $10\%$ size, we take its mean multiplicity from data
and use a corresponding Poisson distribution to sample the multiplicity of each event.
In this ESE analysis, only middle centrality events are used; 
the average multiplicity is $\refm \approx 179$ in the range $-0.5 < \eta < 0.5$.
In short, the default setting of this toy model mimics the Au+Au collisions at $\sqrt{s_{NN}} = 200 \text{ GeV}$.
Total 10.9 billion events are simulated for centrality 20--50\%.



\begin{figure*}
	\includegraphics[width=0.4\linewidth]{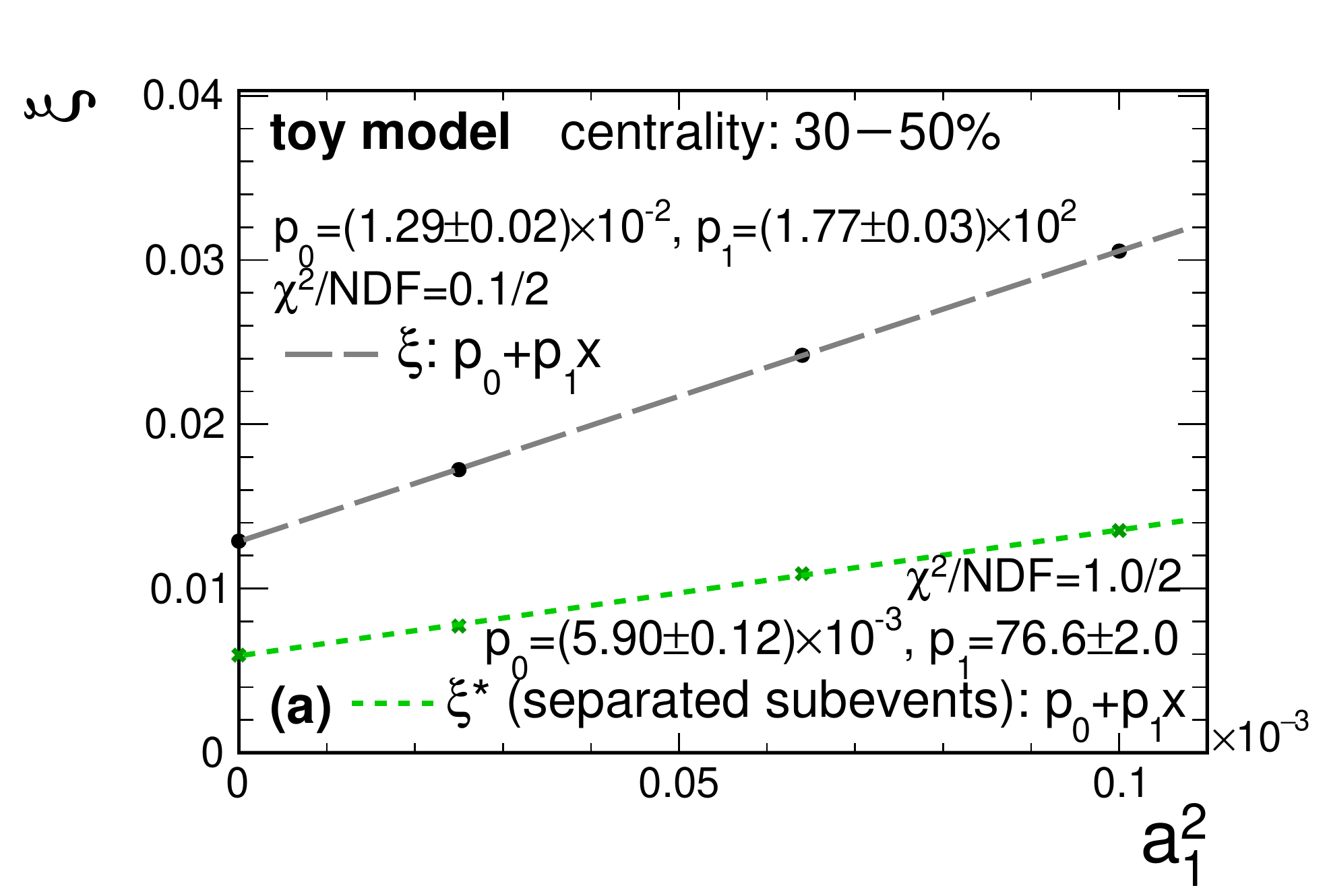}
	\includegraphics[width=0.4\linewidth]{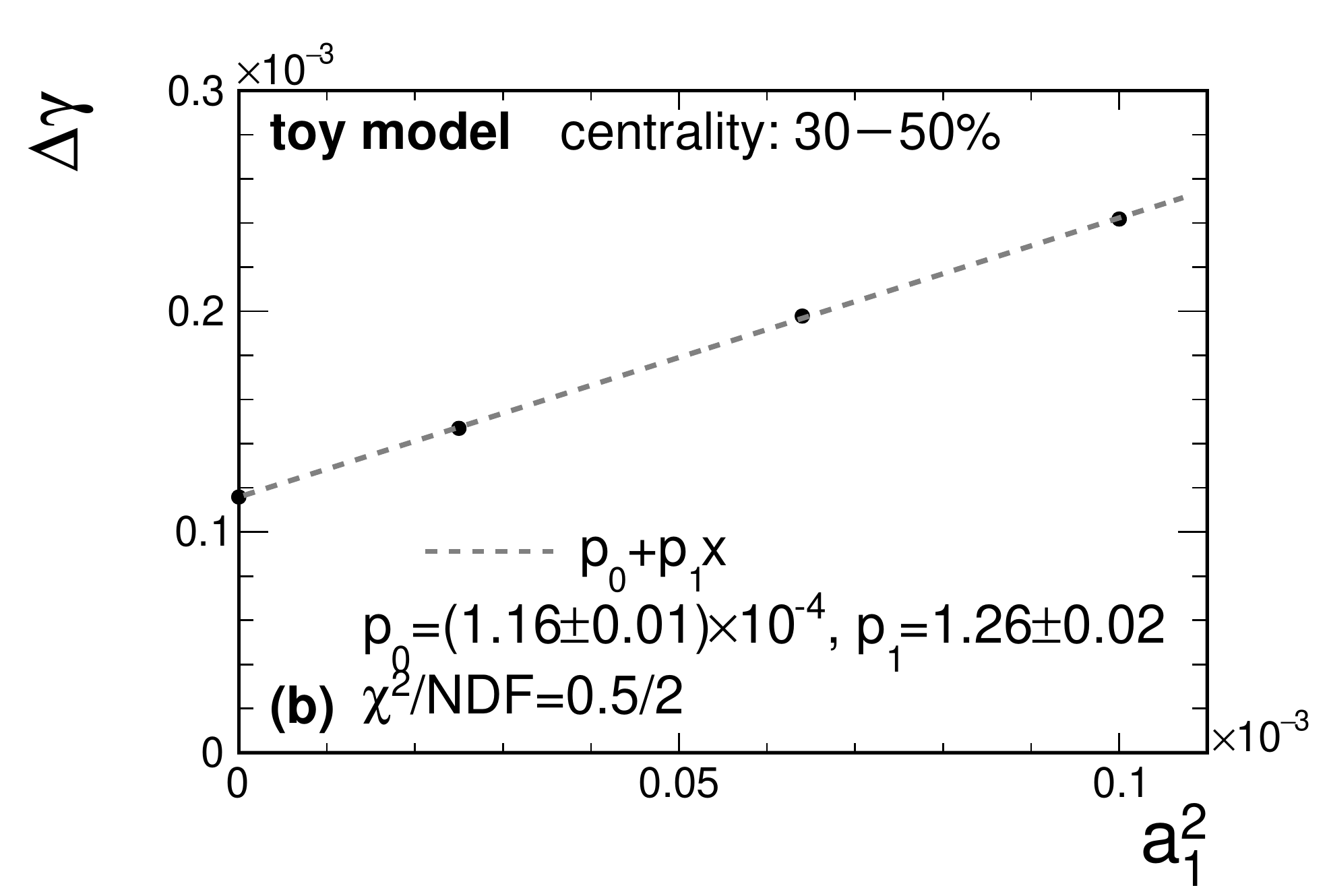}
	\caption{
	The $\kal$ and $\skal$ (a) and $\Delta\gamma$ (b) as functions of the input CME signal ($a_{1}^{2}$) from
	the toy model simulations
	for Au+Au collisions at $\sqrt{s_{NN}}=200 \text{ GeV}$ with default $v_{2}$ distribution.
	Each data point has total 2 billion toy model events in the 0--80\% centrality range
	(or $0.73$ billion events in the 30--50\% centrality range).
	The POI are required to have $0.35\text{ GeV/c} < p_{T} < 2.0 \text{ GeV/c}$ and $0.1< \pm \eta <1.0$,
	whereas particles for EP reconstruction are required to have 
	$0.2 \text{ GeV/c} < p_{T} < 2.0 \text{ GeV/c}$ and $0.1< \mp \eta <1.0$.
	For each dataset,
	the $\kal$, $\skal$, $\Delta\gamma$,
	and $v_{2}$ are calculated from centrality range 30--50\% 
	(average multiplicity $\refm \approx 140$, 
	average subevent POI multiplicity $\mult \approx 64$).
	The dashed lines are two-parameter linear fits.
	}
	\label{ToyAveA1InvWth}
\end{figure*}

\begin{figure*}
	\includegraphics[width=0.325\linewidth]{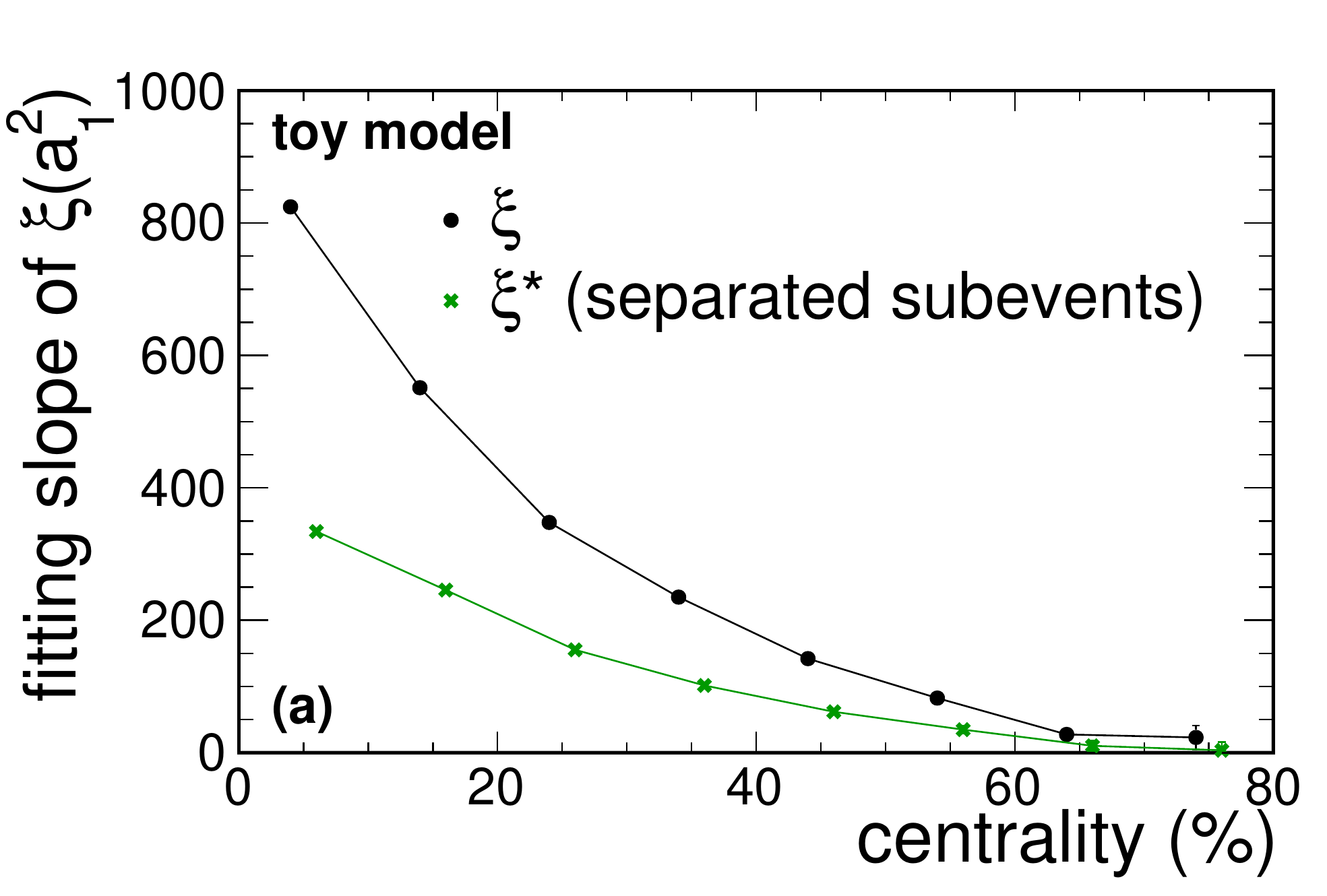}
	\includegraphics[width=0.325\linewidth]{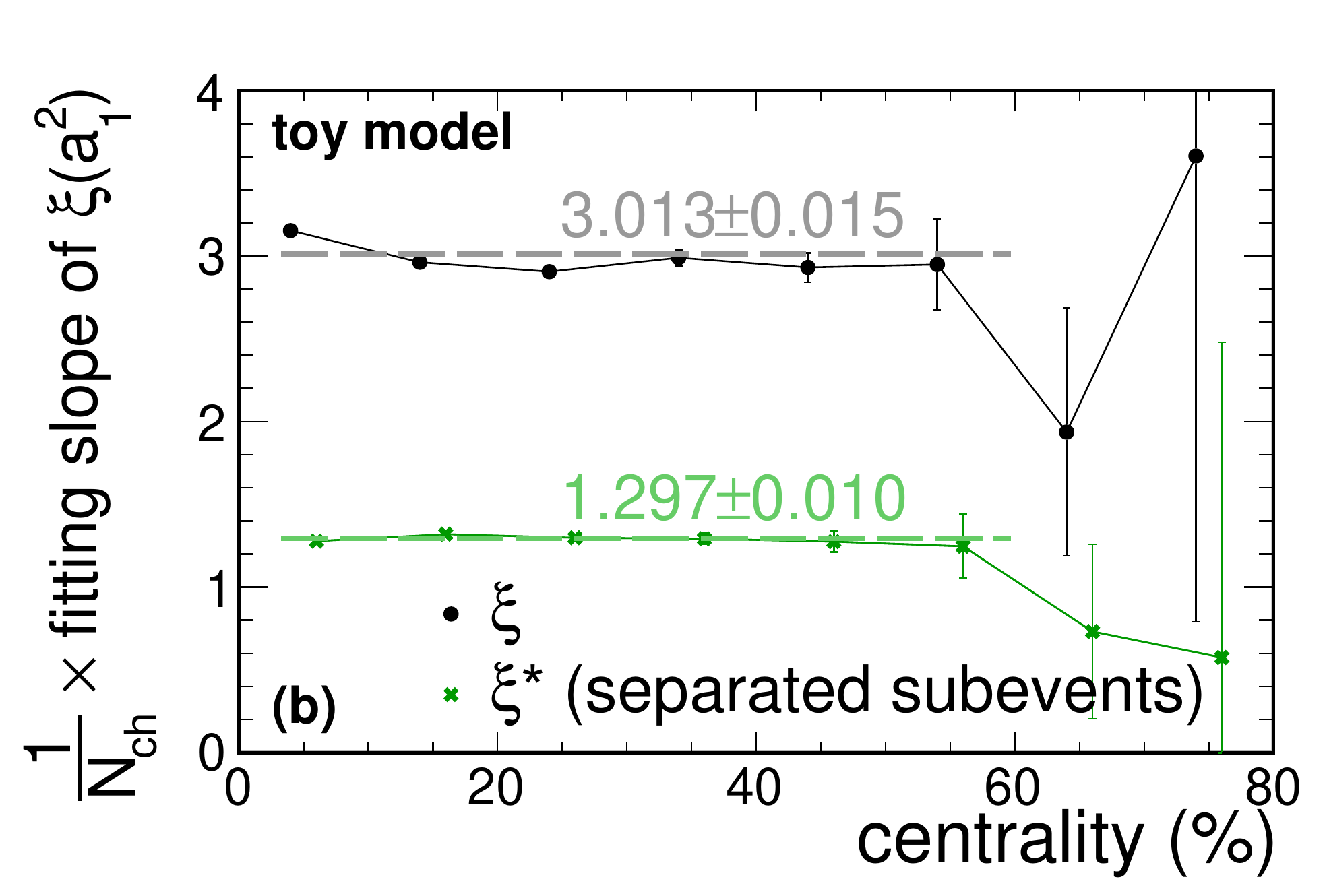}
	\includegraphics[width=0.325\linewidth]{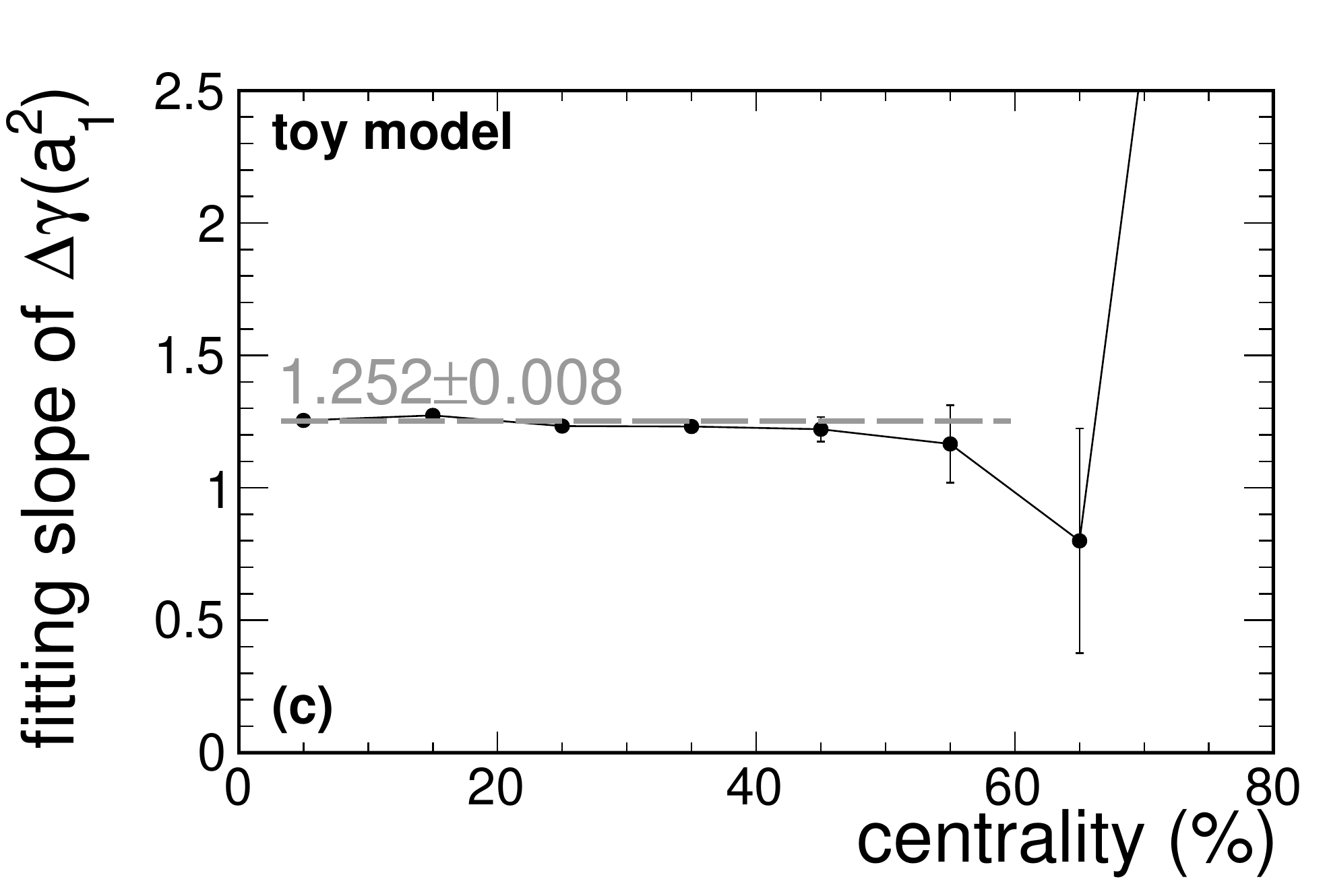}
	\caption{
	The fit parameters from linear fits to $\kal(a_{1}^{2})$, $\skal(a_{1}^{2})$, and $\Delta\gamma(a_{1}^{2})$ 
	are plotted as functions of centrality.
	The $\kal$, $\skal$, $\Delta\gamma$ are similar to those in Fig.~\ref{ToyAveA1InvWth}, 
	but are calculated for each narrow centrality bin of 10\% size.
	(a) The $\kal(a_{1}^{2})$ and $\skal(a_{1}^{2})$ slopes, 
	which are decreasing with increasing centrality percentile (or decreasing centrality).
	(b) The $\kal(a_{1}^{2})$ and $\skal(a_{1}^{2})$ slopes divided by $\mult$.
	The slopes are proportional to $\mult$.
	(c) The $\Delta\gamma(a_{1}^{2})$ slope, which seems to be independent from centrality.
	}
	\label{A1Cent}
\end{figure*}

We followed the STAR analysis by dividing the $|\eta|<1$ particles in the toy model 
into three (east, west, and middle) subevents.
The $q_{2}$ distribution from the middle subevents is shown in Fig.~\ref{Q2DistComp}.
For the $q_{2}$ binning,
equal $q_{2}$ size (except the last $q_{2}$ bin) is taken~\cite{Magdy:2020csm},
which is indicated by the vertical lines in Fig.~\ref{Q2DistComp}.
The five bins are labeled as bin\#1, bin\#2, bin\#3, bin\#4, and bin\#5, respectively,
corresponding to the notations 0--20\%, 20--40\%, 40--60\%, 60--80\%, and 80--100\% in Ref.~\cite{Magdy:2020csm}.
We calculate the average $q_{2}$ values for each ESE $q_{2}$ bin of Fig.~\ref{Q2DistComp} (a).
The elliptic flow of the east and  west subevent are obtained from the two-particle cumulant method,
\begin{equation}
	\vvese = \sqrt{ \langle \cos 2(\phi_{a \in E} - \phi_{b \in W}) \rangle }
	,
\end{equation}
where $a$ is a praticle from east subevent and $b$ from west subevent. 
Figure~\ref{Q2DistComp} (b) shows $\vvese$ as a function of the $q_{2}$. 
The $\vvese$ is found to increase with $q_{2}$, 
indicating some level of selectivity of $v_{2}$ by $q_{2}$.

Figure 4 shows $\kal$ as functions of $q_{2}$ and $\vvese$.
The fits show that $\kal$ incleases with $q_{2}$ and $\vvese$, 
with the slope parameters deviating from zero by approximately 2 standard deviations,
with the current 10.9 billion events simulated for 20-50\% centrality. 
We can make two observations from our ESE study: 
(1) the $\kal$ does depend on $v_{2}$; 
and (2) such ESE studies require humongous statistics in order to draw clear conclusions. 
The latter is probably the primary reason why STAR did not observe a $q_{2}$ dependence of $\kal$ 
with their limited statistics of $\sim 200$ million events for 20-50\% centrality Au+Au collisions~\cite{Magdy:2020csm}.
With the large uncertainties in Ref.~\cite{Magdy:2020csm}, 
it is premature to draw the conclusion 
``$R_{\Psi_{2}}$ is relatively insensitive to $v_{2}$''~\cite{Magdy:2020csm}.


\section{Toy model study to decipher $R_{\Psi_{2}}$} \label{ToyModelResults}

We use the toy model in Sec.~\ref{SecEse} to further study the sensitivity of $R_{\Psi_{2}}$ to background $v_{2}$ and signal $a_{1}$
in Au+Au and isobar collisions.
We simulate all centrality bins measured in data, namely 0--80\%.
To investigate the middle central collisions in greater details,
each 10\%-size bin in 20--50\% (20--60\%) centrality range has twice as many events as other 10\%-size centrality bins in Au+Au (isobar) collisions.

\subsection{Sensitivity to $v_{2}$ background} \label{V2Dependence}

The toy model datasets are simulated with various input $v_{2}(p_{T})$,
including the default $v_{2}$ and variations with 10\%, 20\%, and 30\% increase from the default $v_{2}$.
For each dataset of a given input $v_{2}$,
the $\kal$ and $v_{2}$ of final-state particles are calculated.
Figure~\ref{ToyAveV2InvWth} (a) maps those two variables for the 30--50\% centrality
and a linear dependence is observed between them.
Previous toy model study also observed a
$R_{\Psi_{2}}$ dependence on $v_{2}$ (and transverse momentum $p_{T}$)~\cite{Feng:2018so}.

Similarly, $\Delta\gamma$ is also calculated from those datasets with the same cuts,
and is shown in Fig.~\ref{ToyAveV2InvWth} (b). 
A linear dependence on $v_{2}$ is also observed, 
as one expects for the background behavior in $\Delta\gamma$ as discussed 
in Sec.~\ref{GammaDFDefinition} (cf.~Eq.~\ref{GammaDFv2}).
First-order polynomial fit yields an intercept consistent with zero for $\Delta\gamma(v_{2})$.
This is expected because the $\Delta\gamma$ will go to zero where there is no elliptic flow.

However, first-order polynomial fit to $\kal(v_{2})$ yields a nonzero intercept. 
This is shown in Fig.~\ref{ToyAveV2InvWth} (a). 
The fit parameters are consistent with those from the ESE study shown in Fig.~\ref{SumXEse} (b), 
modulo the large errors for the latter. 
The nonzero intercept arises 
because of the additional correlation between POI and EP 
brought by averaging the two subevents in Eq.~\ref{AveSubDS},
\begin{equation} \label{SelfCorr}
\begin{split}
	&\text{Var}[\Delta S] = \left\langle \Delta S^{2} \right\rangle \\
	=& \frac{1}{4}\left\langle \left(\Delta S^{E}\right)^{2} \right\rangle 
	+  \frac{1}{4}\left\langle \left(\Delta S^{W}\right)^{2} \right\rangle
	+ \frac{1}{2} \left\langle \Delta S^{E} \Delta S^{W} \right\rangle
	.
\end{split}
\end{equation}
This auto-correlation comes about 
because the POI for $\Delta S^{E}$ are used for EP reconstruction for $\Delta S^{W}$, and vice versa. 
To circumvent this, we count the two subevents separately instead of combining them.
The squared inverse width of $R_{\Psi_{2}}(\Delta S'')$ obtained from this method, is referred to as $\skal$.
The $\skal(v_{2})$ is shown in Fig.~\ref{ToyAveV2InvWth} (a). 
A linear dependence is observed, with an intercept consistent with zero. 
In fact, the difference between $\kal$ and $\skal$ at any given setting (i.e.~not just the intercept as we noted above) is caused by the auto-correlation. This will be discussed further in Appendix~\ref{EPres}.\ref{EPresSTAR} and \ref{EPres}.\ref{SelfKal}

We repeat the fit for each narrow centrality bin.
In Fig.~\ref{V2Cent} (a),
the fit intercept parameter is shown as a function of centrality for $\kal(v_{2})$,
which seems roughly a constant in the centrality range 10--50\%.
In Fig.~\ref{V2Cent} (b), 
the fit slope parameter is shown as functions of centrality for $\kal(v_{2})$, $\skal(v_{2})$.
The $\kal(v_{2})$ ($\skal(v_{2})$) slope is roughly constant in the centrality range 10--50\% with a value approximately 0.105 (0.098).
See Appendix~\ref{CalcXi}.\ref{BkgV2Dep} for an analytical derivation.

Figure~\ref{V2Cent} (c) shows the slope of $\Delta\gamma(v_{2})$ multiplied by $\mult$,
where $\mult$ is average POI multiplicity of each subevent 
in $-1.0<\eta<-0.1$ (or $0.1<\eta<1.0$) with $p_{T}>0.35 \text{ GeV/c}$.
It is found that 
the slope parameter of $\Delta\gamma(v_{2})$ is inversely proportional to multiplicity, 
with a dependence approximately $0.125/\mult$.
This is consistent with previous findings that the $\Delta\gamma$ is diluted by multiplicity, 
as discussed in Sec.~\ref{GammaDFDefinition} (cf.~\ref{GammaDFv2}).
See further details in Appendix~\ref{CalcDg}.

To summarize, the $\kal$, $\skal$, and $\Delta\gamma$ can be parameterized, 
empirically for the given toy model simulation in this study without CME signal input, as
\begin{subequations} \label{TMv2Dep}
\begin{align}
	\kal_{\rm bkgd} \approx& (0.105\pm0.011) v_{2} +(0.0052\pm0.0007), \\
	\skal_{\rm bkgd}  \approx& (0.098 \pm 0.007) v_{2},  \\
	\Delta\gamma_{\rm bkgd} \approx& (0.125 \pm 0.006) v_{2} / \mult . 
\end{align}
\end{subequations}

\subsection{Responses to CME signal} \label{A1Dependence}

To study the sensitivity to CME signal, 
we input an $a_{1}$ parameter into particle distribution in the toy model, 
keeping the default setting for the $v_{2}$ background.
We set $a_{1}$ to $0$, $0.005$, $0.008$, and $0.010$.
For each case, we generate 2 billion events over 0--80\% centrality,
where 0.73 billion events are in 30--50\% centrality.

The $\kal$, $\skal$, and $\Delta\gamma$ are calculated
from centrality range 30--50\%.
Figure~\ref{ToyAveA1InvWth} (a) and (b) show 
the $\kal$, $\skal$ and $\Delta\gamma$ as functions of $a_{1}^{2}$. 
Linear dependence on $a_{1}^{2}$ is observed for all observables. 
Linear fits are superimposed in Fig.~\ref{ToyAveA1InvWth}~(a,b).
All show nonzero intercepts,
corresponding to the backgrounds caused by the nonzero $v_{2}$ in the underlying events.

The similar procedure above is then repeated for each narrow centrality bin.
Figure~\ref{A1Cent} (a) shows
the fit slope parameters as functions of centrality 
for $\kal(a_{1}^{2})$, $\skal(a_{1}^{2})$.
The $\kal(a_{1}^{2})$ slope is found to decrease with centrality percentile (or increase with centrality); 
it is found to be proportional to multiplicity (Fig.~\ref{A1Cent} (b)).
See Appendix~\ref{CalcXi}.\ref{SigA1Dep} for an analytical derivation.

Figure~\ref{A1Cent} (c) shows 
the fit slope parameters as a function of centrality for $\Delta\gamma(a_{1}^{2})$.
The $\Delta\gamma(a_{1}^{2})$ slope parameter is found to be independent of the centrality,
and intercept is always consistent with zero.
Ideally one expect $\Delta\gamma$ to vary as $2 a_{1}^{2}$ (Eq.~\ref{GammaDFa1}). 
The slope parameter from our toy model is found to be $\sim 1.252$, smaller than $2$; 
this is because the CME signal $a_{1}$ is applied only to primordial pions, 
not to the secondary pions from resonance decays.
See further details in Appendix~\ref{CalcDg}. 
If parameter $a_{1}$ characterizes the coefficient in Eq.~\ref{PrimoDist} which includes all final-state particles, then we would have $\Delta\gamma=2 a_{1}^{2}$. 

To summarize, the CME signal dependence of the $\kal$, $\skal$, and $\Delta\gamma$ can be parameterized,
empirically for the given toy model simulation in this study, as
\begin{subequations} \label{TMa1Dep}
\begin{align}
	\kal_{\rm CME} \approx& (3.013 \pm 0.015) a_{1}^{2} \mult , \\
	\skal_{\rm CME} \approx & (1.297 \pm 0.010) a_{1}^{2} \mult, \\
	\Delta\gamma_{\rm CME} \approx& (1.252 \pm 0.008) a_{1}^{2} .
\end{align}
\end{subequations}

\subsection{Relative merits of $R_{\Psi_{2}}$ and $\Delta\gamma$}

\begin{figure}
	\includegraphics[width=1.0\linewidth]{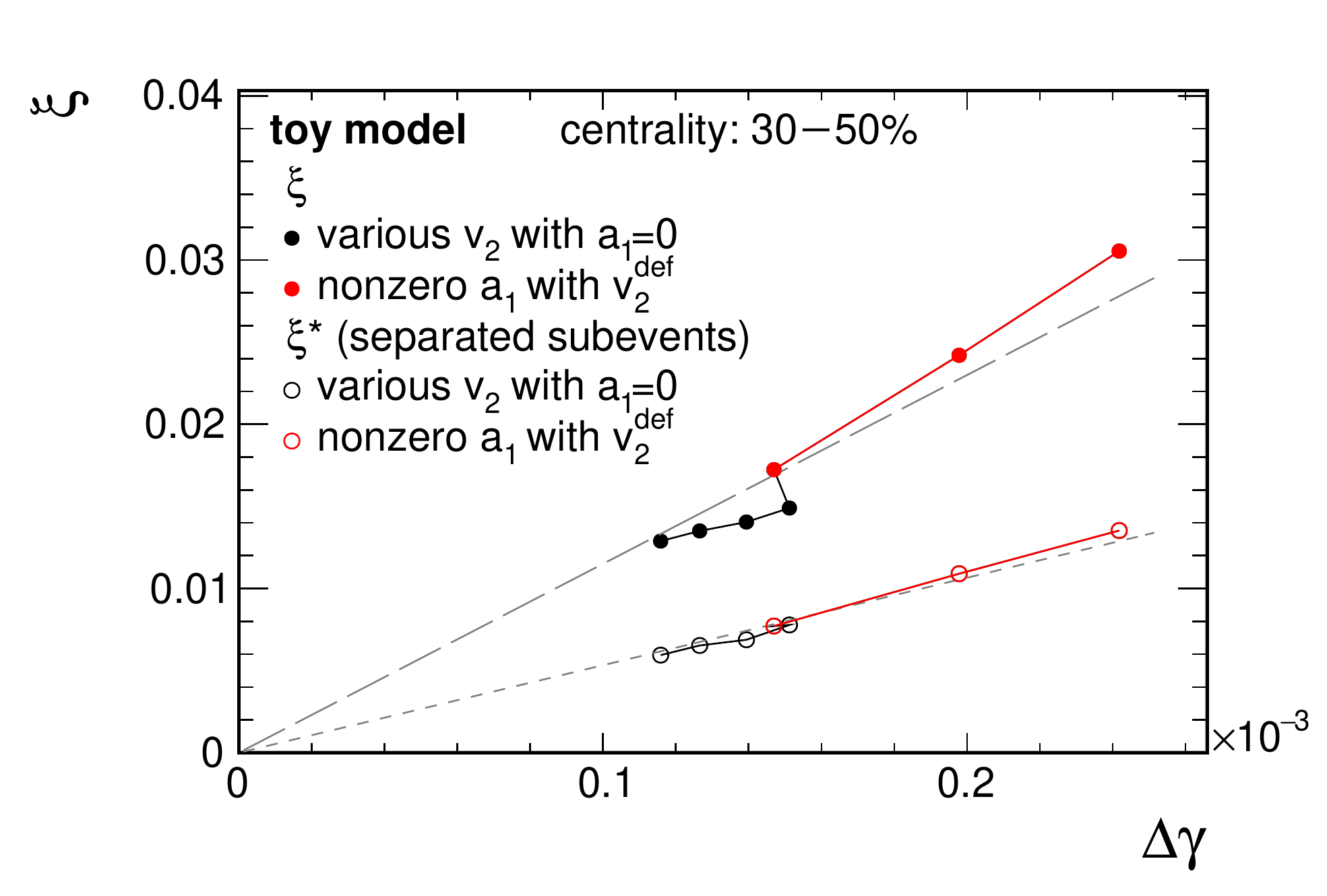}
	\caption{
	The $\kal$ ($\skal$) and $\Delta\gamma$
	are mapped for each toy model simulation for Au+Au collisions at $\sqrt{s_{NN}}=200 \text{ GeV}$
	in Fig.~\ref{ToyAveV2InvWth}
	(various input $v_{2}$ distributions with no CME signal, first four data points)
	and Fig.~\ref{ToyAveA1InvWth}
	(various input $a_{1}$ values with default $v_{2}$ distribution, last three data points).
	The lines are one-parameter linear fits to the data points, mainly to guide the eye.
	}
	\label{ToyAveGammaDFKal}
\end{figure}

To summarize the findings in Sec.~\ref{V2Dependence} and \ref{A1Dependence}, 
we can parameterize $\kal$, $\skal$, and $\Delta\gamma$ 
in terms of the $v_{2}$ background and the CME signal, 
in our toy model simulation of Au+Au 10--50\% centrality, by:
\begin{subequations} \label{TMa1v2Dep}
\begin{align}
	\kal/ \mult \approx& (0.105\pm0.011) v_{2} /\mult + (3.013\pm0.015) a_{1}^{2}  \nonumber \\
	&  + (0.0052\pm0.0007)/\mult, \\
	\skal/ \mult \approx& (0.098\pm0.007) v_{2} / \mult + (1.297 \pm 0.010) a_{1}^{2}, \\
	\Delta\gamma \approx& (0.125 \pm 0.006) v_{2} /\mult + (1.252\pm0.008) a_{1}^{2} 
	.
\end{align}
\end{subequations}
It is worthwhile to note that $\kal/\mult$, and specially $\skal/\mult$, is rather similar to $\Delta\gamma$. 
This may not be surprising as the $\kal$ is related to the combination of the $\Delta S$ and $\Delta S^{\perp}$ variances, roughtly 
$\langle \cos(\phi_{a} - \rp) \cos(\phi_{b} - \rp)- \sin(\phi_{a} - \rp)\sin(\phi_{b} - \rp) \rangle$, 
which is the $\Delta\gamma$~\cite{Voloshin:2017drupal}.
In Appendix~\ref{CalcXi}, we provide an analytical derivation of $\skal$ 
without considering correlations arising from $p_{T}$ dependence of $v_{2}$ and decay kinematics, etc. 
Our toy model results above and the analytical estimates are qualitatively consistent.
Our analytical results are also qualitatively in line with findings by others~\cite{Tang:2020drupal}.

We may estimate the signal/background ratio ($S/B$) of the two observables from Eq.~\ref{TMa1v2Dep}, within our toy model simulation, as
\begin{subequations} \label{TMa1v2SB}
\begin{align}
	\kal:& \quad S/B \approx \frac{ (28.7 \pm 3.0) a_{1}^{2} \mult }{ v_{2} +(0.050\pm0.008)}, \\
	\skal: & \quad S/B \approx (13.2 \pm 1.0) a_{1}^{2} \mult / v_{2} \\
	\Delta\gamma:& \quad S/B \approx (10.0 \pm 0.5) a_{1}^{2} \mult / v_{2}
	.
\end{align}
\end{subequations}
Thus, in terms of the $S/B$ value,
$\skal$ is more (less) sensitive to signal (background) than $\Delta\gamma$
with this toy model in this centrality range.

We can map the observables $\kal$ vs.~$\Delta\gamma$ and $\skal$ vs.~$\Delta\gamma$ 
against each other using the data in
Sec.~\ref{V2Dependence} and \ref{A1Dependence}, as shown in Fig.~\ref{ToyAveGammaDFKal}.
There is a monotonic, one-to-one correspondence between $\skal$ and $\Delta\gamma$,
indicating that they are essentially equivalent in searching for the CME. 
A recent  AMPT simulation study also shows that the $R_{\Psi_{2}}$ and $\Delta\gamma$ observables are essentially equivalent~\cite{Yao:2020dnp}.
For $\kal$ and $\Delta\gamma$, there are two groups of data points with different slopes, 
one from background variation and the other from signal variation. 
This is likely caused by the auto-correlations arising from averaging $\Delta S$ between subevents discussed in Sec.~\ref{V2Dependence}  and Appendix~\ref{EPres}. 
Our toy model study only includes $\rho$ decays, 
while the real collisions have also other resonances whose decay kinematics are different from the $\rho$'s. 
This can render possible quantitative changes in the relative merits of $\Delta\gamma$ and $\kal$, $\skal$.

It is worthwhile to note, however, that the $\Delta\gamma$ observable is relatively straightforward to interpret
whereas the $R_{\Psi_m}$ observabale is complex. 
The $\Delta\gamma$ variable is computed per particle pair and the $\kal$ ($\skal$) variable is computed per event.
The former offers a wider versatility in ways to isolate the CME signal from backgrounds, 
for example, a differential study in pair invariant mass~\cite{Zhao:2017nfq, Zhao:2017wck, Li:2018oot, Zhao:2018blc}. 
Although $\kal$ ($\skal$) may have a slightly larger $S/B$ value 
than $\Delta\gamma$ according to our toy model study, 
both are strongly affected by physics backgrounds which dominate over the CME. 
Both observables have to seek innovative ways to isolate the CME signal and physics backgrounds. 
One of the promising ways is to leverage on the different harmonic planes in the same collision event for $\Delta\gamma$ measurements~\cite{Xu:2018cpc, Xu:2018wvm, Zhao:2020utk}. 
It would be interesting to study the benefit of applying such a method to $\kal$ ($\skal$).


%

%
\subsection{Isobar background expectations}

Recently, $^{96}_{44}\text{Ru}$+$^{96}_{44}\text{Ru}$,
and $^{96}_{40}\text{Zr}$+$^{96}_{40}\text{Zr}$
collisions have been conducted at RHIC to potentially resolve the background issue in the search for CME. 
Those two species are isobars of each other, 
with the same number of nucleons ($A = 96$) but different number of protons ($Z = 44, 40$).
The backgrounds are expected to be similiar in those two collision systems due to the same nucleon number.
The CME signals should be quite different due to the different magnetic fields 
created by the spectator protons whose numbers are different in those isobars.
There may be complications to these simple expectations 
when considering modern nuclear structure calculations~\cite{Xu:2018prl, Xu:2018wvm}.

Both the $R_{\Psi_{m}}$ and the inclusive $\Delta\gamma$ correlators are employed to search for the CME using the isobar data~\cite{Magdy:2018iso}.
To examine the relative merits of the two observables in searching for the CME, 
we simulate Ru+Ru and Zr+Zr collisions using the toy model.
The multiplicity is scaled from Au+Au by the number of participant nucleons.
We use the following inputs for the Zr+Zr and Ru+Ru collision systems, respectively.
\begin{itemize}
	\item Zr+Zr: default $v_{2,\pi}(p_{T})$ and $v_{2,\rho}(p_{T})$ 
	(as same as those used in the default Au+Au toy model simulation), and the CME signal $a_1=0.50\%$;
	\item Ru+Ru: 2\% larger $v_{2,\pi}$ and $v_{2,\rho}$, and a 10\% larger CME signal than in Zr+Zr.
	The 2\% larger $v_{2}$ is guided by the expected $v_{2}$ difference from modern nuclear structure calculations~\cite{Xu:2018prl, Xu:2018wvm}. 
	The 10\% larger CME signal comes from the 10\% more protons in the Ru than Zr nucleus.
\end{itemize}

\begin{figure}
	\includegraphics[width=1.0\linewidth]{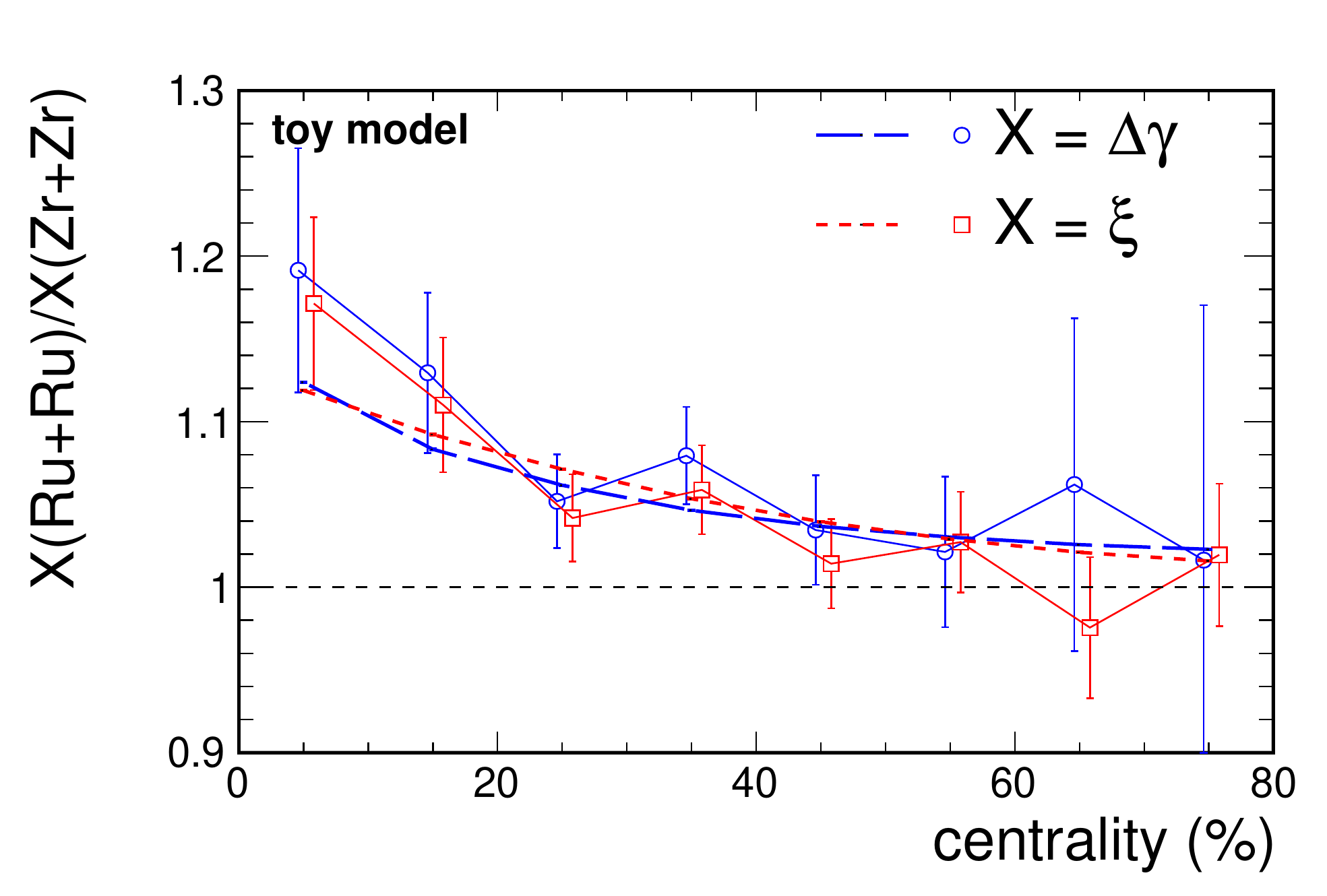}
	\caption{The ratio between the observable $X$ in the two collision systems as functions of the centrality,
	where $X$ is $\kal$ or $\Delta\gamma$.
	For Zr+Zr, the default $v_{2}^{\text{def}}$ and CME signal $a_{1} = 0.005$ are input,
	whereas for Ru+Ru, the $1.02 v_{2}^{\text{def}}$ and $a_{1} = 0.0055$ are input.
	Each dataset has 3 billion toy model events in the 0--80\% centrality range,
	where 2 billion events are in 20--60\% centrality range.
	The POI are required to have $0.35\text{ GeV/c} < p_{T} < 2.0 \text{ GeV/c}$ and $0.1< \pm \eta <1.0$,
	whereas particles for EP reconstruction are required to have 
	$0.2 \text{ GeV/c} < p_{T} < 2.0 \text{ GeV/c}$ and $0.1< \mp \eta <1.0$.
	The curves are given by Eqs.~\ref{IsoGammaDF} and \ref{IsoR2Sigma}.
	}
	\label{Cme10CompRtCent}
\end{figure}

Figure~\ref{Cme10CompRtCent} shows the Ru+Ru over Zr+Zr ratio of $\kal$ in the two isobar systems, 
along with that of $\Delta\gamma$.
The centrality dependence of the ratios can be understood by Eq.~\ref{TMa1v2Dep}.
For $\Delta\gamma$, the ratio is 
\begin{equation} \label{IsoGammaDF}
\begin{split}
	&\frac{\Delta\gamma (\text{Ru+Ru}) }{\Delta\gamma (\text{Zr+Zr})} \\
	=& \frac{ 1.252 a_{1}^{2}\times1.1^2 + 0.125 v_2/ \mult \times1.02 }{ 1.252 a_{1}^{2} + 0.125 v_2/ \mult }\\
	=& 1.21 - \frac{0.19\times0.125 v_2}{ 1.252 a_{1}^{2} \mult + 0.125 v_2 } \\
	=& 1.02 + \frac{0.19\times1.252 a_{1}^{2} \mult }{ 1.252 a_{1}^{2} \mult + 0.125v_2 }
	,
\end{split}
\end{equation}
and the double ratio is
\begin{equation} \label{IsoGammaDFV2}
\begin{split}
	&\frac{\Delta\gamma (\text{Ru+Ru}) / v_{2}(\text{Ru+Ru})}{\Delta\gamma(\text{Zr+Zr}) / v_{2}(\text{Zr+Zr})} \\
	=& \frac{\Delta\gamma (\text{Ru+Ru}) }{\Delta\gamma (\text{Zr+Zr})} \times \frac{1}{1.02} \\
	=& 1.186 - \frac{0.186\times0.125 v_2}{ 1.252 a_{1}^{2} \mult + 0.125 v_2 } \\
	=& 1 + \frac{0.186\times1.252 a_{1}^{2} \mult }{ 1.252 a_{1}^{2} \mult + 0.125v_2 }
	.
\end{split}
\end{equation}
For $\kal$, the ratio is 
\begin{equation} \label{IsoR2Sigma}
\begin{split}
	& \frac{\kal (\text{Ru+Ru}) }{\kal (\text{Zr+Zr}) } \\
	=& \frac{ 3.013 a_{1}^{2}\times1.1^2 + 0.105 v_2/ \mult \times1.02 + 0.0052 /\mult}{ 3.013 a_{1}^{2} + 0.105 v_2/ \mult + 0.0052 /\mult }\\
	=& 1.21 - \frac{0.19\times0.105 v_2 + 0.21\times 0.0052 }{ 3.013 a_{1}^{2} \mult + 0.105 v_2 + 0.0052} \\
	=& 1.02 + \frac{0.19\times 3.013 a_{1}^{2} \mult - 0.02\times0.0052 }{ 3.013 a_{1}^{2} \mult + 0.105 v_2 + 0.0052}
	.
\end{split}
\end{equation}
The last two lines of Eqs.~\ref{IsoGammaDF}, \ref{IsoGammaDFV2}, and  \ref{IsoR2Sigma} are different ways to express the ratios to illustrate the limits.
In the limit of high multiplicity $\mult \rightarrow \infty$ (centrality $\rightarrow 0\%$), the ratios go to 1.21,
and the double ratio goes to $1.186$.
In the limit of low multiplicity $\mult \rightarrow 0$ (centrality $\rightarrow 80\%$), the ratios go to 1.02 ($\Delta\gamma$) or $\sim1.01$ ($\kal$), and the double ratio goes to $1$.
We superimpose in Fig.~\ref{Cme10CompRtCent} the parameterizations of Eqs.~\ref{IsoGammaDF} and \ref{IsoR2Sigma}.
Since the trends and statistic errors are similar for $\kal$ and the inclusive $\Delta\gamma$, as evident from Fig.~\ref{Cme10CompRtCent}, the two observables would serve the same functionality in searching for the CME in isobar collisions; neither seems superior to the other. 
The conclusion is the same if $\skal$ is used instead of $\kal$.

It is worthwhile to note that our toy model simulation is useful and informative to reveal the relative merits of the $R_{\Psi_{2}}$ and $\Delta\gamma$ observables within the same simulated data. 
One, however, should not take the magnitudes and error bars of the points in Fig.~\ref{Cme10CompRtCent} to infer those of the real isobar data. 
Even though we simulated the similar number of events as in data, 
the physics included in our toy model is overly simplified (e.g., only $\rho$ resonance is included) 
and the CME signal is, of course, an arbitrary input.


\section{Summary} \label{Summary}



We have studied the $R_{\Psi_{m}}$ correlators using the AMPT model, which does not include any CME signal.
With Au+Au collisions at $\sqrt{s_{NN}} = 200 \text{ GeV}$ simulated by AMPT, the $R_{\Psi_{2}}$ distribution is concave. 
The $R_{\Psi_{3}}$ distribution, with the choice of the azimuthal angle range of $\phi\in[-\pi,\pi)$, is concave and differs from that of $R_{\Psi_{2}}$, but with the choice $\phi\in[0,2\pi)$, it is approximately flat indicating the illness of the $R_{\Psi_{3}}$ definition~\cite{Feng:2018so}.
The same AMPT model is also used to simulate small-system p+Au and d+Au collisions at $\sqrt{s_{NN}} = 200 \text{ GeV}$. The $R_{\Psi_{2}}$ distributions are found to be slightly concave in those small-system collisions.

We have used a toy model to generate primordial and resonance decay pions, according to kinematic distributions and elliptic flow measured in 200 GeV Au+Au collision data. 
It is found that the $R_{\Psi_{m}}$ distribution squared inverse width ($\skal$) is proportional to $v_2$. We verified the approximate linearity with algebraic derivation.
In addition, we find that the usual implementation of $R_{\Psi_{2}}$ by averaging subevents 
introduces an auto-correlation that causes an intercept in the $\kal(v_{2})$ linear dependence.
We have also input CME signal into the toy model via the $a_1$ parameter. 
It is found that the $\kal$ and $\skal$ increase linearly with $\mult a_1^2$, 
where $\mult$ is the multiplicity of the particles of interests.
We have also calculated the $\Delta\gamma$ correlator and found the expected linear dependence on $v_2/ \mult$ and on $a_1^2$.
Except the multiplicative factor of $\mult$, the dependences on $v_2$ and $a_1^2$ are rather similar between $\skal$ and $\Delta\gamma$,
and also between $\kal$ and $\Delta\gamma$.

The toy model simulation, with only $v_2$ background, is also used for an event shape engineering study. 
It is found that the $\kal$ does depend on the event-by-event $q_2$ and $v_2$
at 2 sigma significance with 10.9 billion events corresponding to 20--50\% centrality Au+Au collisions. 

The toy model is also used to simulate the isobar systems at $\sqrt{s_{NN}} = 200 \text{ GeV}$.
With the anticipated 10\% CME signal ($a_1$) and 2\% flow background ($v_2$) differences, 
the $\kal$ and the inclusive $\Delta\gamma$ relative strengths between the isobar collision systems 
have rather similar trend on centrality, with similar magnitudes and statistical uncertainties.
It appears that the two observables are essentially the same; 
neither observable has advantage over the other.

It has been argued~\cite{Magdy:2020csm} that 
(i) the $R_{\Psi_{2}}$ and $R_{\Psi_{3}}$ distributions were identical for pure background scenarios, 
(ii) the small-system collisions yield flat $R_{\Psi_{2}}$ distributions, 
and (iii) the $R_{\Psi_{2}}$ distribution does not depend on $q_2$ 
with event shape engineering where variation in $v_2$ is observed. 
These corroborative features led to the conclusion that the concave $R_{\Psi_{2}}$ distribution observed in Au+Au collisions, more strongly concave than the $R_{\Psi_{3}}$ distribution, is inconsistent with known backgrounds and thus may suggest the presence of the CME signal~\cite{Magdy:2020csm}. 
Our studies indicate that none of the three features seems to uphold, 
and there appears to be no qualitative difference between the $R_{\Psi_{2}}$ observable and the inclusive $\Delta\gamma$ correlator.


\section*{Acknowledgments}

This work is supported in part by the U.S.~Department of Energy Grant No.~DE-SC0012910 
and the National Natural Science Foundation of China Grant Nos.~11905059, 12035006, 12047568, 12075085.


\appendix

\section{EP resolution corrections} \label{EPres}

In this appendix, we first derive the analytical form of the EP resolution correction factor (Eq.~\ref{eq:rm1}) for the squared inverse width of the $R_{\Psi_m}$ correlator, $\skal$. We then discuss the empirical correction factor used by STAR~\cite{Magdy:2020csm}. Finally we investigate the effect of auto-correlations on $\kal$.

\subsection{EP resolution correction for $\skal$} \label{EPresCalc}

Ideally, one likes to use the RP $\Psi_{\text{RP}}$ in Eq.~\ref{SubDS} 
instead of subevent EP $\Psi_{m}$.
In this section, we derive the correction factor on $\Delta S$ to take into account the inaccuracy of the reconstructed EP in representing the RP. To lighten notations, we do not explicitly specify the subevent by the superscript $E/W$, but rather implicitly refer to a given subevent for the $\Delta S_{m}$ and $\Psi_{m}$ quantities.

The terms in $\Delta S_{m}$ can be written, taking one representative term as an example, into
\begin{equation}
\begin{split}
	&\sin\left(\frac{m}{2} (\phi^{+} - \Psi_{m})\right) \\
	=& \sin\left(\frac{m}{2}(\phi^{+} - \Psi_{\text{RP}})\right)
	\cos\left(\frac{m}{2}(\Psi_{m} - \Psi_{\text{RP}})\right) - \\
	& \cos\left(\frac{m}{2}(\phi^{+} - \Psi_{\text{RP}})\right)
	\sin\left(\frac{m}{2}(\Psi_{m} - \Psi_{\text{RP}})\right)
	.
\end{split}
\end{equation}
Thus, the relationship between the $\Delta S_{m}$ variables w.r.t.~$\Psi_{\text{RP}}$ and $\Psi_{m}$ are
\begin{equation}
\begin{split}
	\Delta S_{m} (\Psi_{m}) 
	=& \Delta S_{m} (\Psi_{\text{RP}}) \cos\left(\frac{m}{2}(\Psi_{m} - \Psi_{\text{RP}})\right) - \\
	& \Delta S_{m}^{\perp} (\Psi_{\text{RP}}) \sin\left(\frac{m}{2}(\Psi_{m} - \Psi_{\text{RP}})\right), \\
	\Delta S_{m}^{\perp} (\Psi_{m}) 
	=& \Delta S_{m}^{\perp} (\Psi_{\text{RP}}) \cos\left(\frac{m}{2}(\Psi_{m} - \Psi_{\text{RP}})\right) +\\
	& \Delta S_{m} (\Psi_{\text{RP}}) \sin\left(\frac{m}{2}(\Psi_{m} - \Psi_{\text{RP}})\right)
	.
\end{split}
\end{equation}
The relationship among the variances (corresponding to the squared widths of the $\Delta S$ distributions) are then
\begin{equation} \label{PPsigma2}
\begin{split}
	\sigma^2 \Big[ \Delta S_{m} (\Psi_{m}) \Big] 
	=& \sigma^2 \Big[ \Delta S_{m} (\Psi_{\text{RP}}) \Big] \frac{1+r_{m}}{2} +\\
	& \sigma^2 \Big[ \Delta S_{m}^{\perp} (\Psi_{\text{RP}}) \Big] \frac{1-r_{m}}{2}, \\
	\sigma^2 \Big[ \Delta S_{m}^{\perp} (\Psi_{m}) \Big] 
	=& \sigma^2 \Big[ \Delta S_{m}^{\perp} (\Psi_{\text{RP}}) \Big] \frac{1+r_{m}}{2} +\\
	& \sigma^2 \Big[ \Delta S_{m} (\Psi_{\text{RP}}) \Big] \frac{1-r_{m}}{2}
	,
\end{split}
\end{equation}
where $r_{m}$ is the resolution of the subevent EP (Eq.~\ref{SubEpRes}).
For convenience, we denote the variances with respect to the RP by
\begin{equation}
\begin{split}
	\realpara =& \sigma \Big[ \Delta S_{m} (\Psi_{\text{RP}}) \Big], \quad\quad
	\realperp = \sigma \Big[ \Delta S_{m}^{\perp} (\Psi_{\text{RP}}) \Big], \\
	\shffpara =& \sigma \Big[ \Delta S_{m,\text{sh}} (\Psi_{\text{RP}}) \Big], \quad
	\shffperp = \sigma \Big[ \Delta S_{m,\text{sh}}^{\perp} (\Psi_{\text{RP}}) \Big]
	.
\end{split}
\end{equation}
Then Eq.~\ref{PPsigma2} becomes
\begin{equation} \label{sigma2abr}
\begin{split}
	\sigma^2 \Big[ \Delta S_{m} (\Psi_{m}) \Big] 
	=& \frac{\realpara^2 + \realperp^2}{2} + \frac{\realpara^2 - \realperp^2}{2} r_{m}, \\
	\sigma^2 \Big[ \Delta S_{m}^{\perp} (\Psi_{m}) \Big] 
	=& \frac{\realpara^2 + \realperp^2}{2} - \frac{\realpara^2 - \realperp^2}{2} r_{m}
	.
\end{split}
\end{equation}

For convenience of presentation, we write in the following the distributions in $\Delta S_{m}$, $\Delta S_{m,\text{sh}}$, $\Delta S_{m}^{\perp}$, and $\Delta S_{m,\text{sh}}^{\perp}$ all as Gaussians, with vanishing means, and variances $\sigma_{m}$, $\sigma_{m,\text{sh}}$, $\sigma_{m}^{\perp}$, and $\sigma_{m,\text{sh}}^{\perp}$. However, our end conclusion is general, independent of whether those distributions are Gaussians or not.
Take the $\Delta S_{m}$ distribution as
\begin{equation}
	\frac{\der N_{\text{event}}}{N_{\text{event}}} = 
	\frac{1}{\sqrt{2\pi\sigma_{m}^{2}}} e^{-\frac{\Delta S_{m}^{2}}{2\sigma_{m}^{2}}} \der(\Delta S_{m})
	.
\end{equation}
The scaled $\Delta S''$ distribution is 
\begin{equation}
	\frac{\der N_{\text{event}}}{N_{\text{event}}} =  \frac{\sigma_{m,\text{sh}}}{\delta_{r_{m}}\sqrt{2\pi\sigma_{m}^{2}}} e^{-\frac{\Delta {S''}_{m}^{2}}{2\delta_{r_{m}}^{2}}\frac{\sigma_{m,\text{sh}}^{2}}{\sigma_{m}^{2}}} \der(\Delta {S''}_{m})
	.
\end{equation}
The shape of the scaled $C_{\Psi_{m}}$ distribution is then 
\begin{equation}
	C_{\Psi_{m}}(\Delta {S''}_{m}) = \frac{\sigma_{m,\text{sh}}}{\sigma_{m}} e^{-\frac{\Delta {S''}_{m}^{2}}{2\delta_{r_{m}}^{2}} \left(\frac{\sigma_{m,\text{sh}}^{2}}{\sigma_{m}^{2}}-1 \right)}
	.
\end{equation}
Similarly, we get
\begin{equation}
	C_{\Psi_{m}}^{\perp}(\Delta {S''}_{m}) = \frac{\sigma_{m,\text{sh}}^{\perp}}{\sigma_{m}^{\perp}} e^{-\frac{\Delta {S''}_{m}^{2}}{2\delta_{r_{m}}^{2}} \left(\frac{{\sigma_{m,\text{sh}}^{\perp}}^{2}}{{\sigma_{m}^{\perp}}^{2}}-1 \right)}
	.
\end{equation}
Finally the shape of the scaled $R_{\Psi_{m}}$ correlator can be written as
\begin{equation} \label{RShape}
\begin{split}
	R_{\Psi_{m}}(\Delta {S''}_{m}) 
	=& \frac{C_{\Psi_{m}}(\Delta {S''}_{m})}{C_{\Psi_{m}}^{\perp}(\Delta {S''}_{m})} \\
	=& \frac{\sigma_{m,\text{sh}}\sigma_{m}^{\perp}}{\sigma_{m,\text{sh}}^{\perp}\sigma_{m}} 
	e^{-\frac{\Delta {S''}_{m}^{2}}{2\delta_{r_{m}}^{2}} 
	\left(\frac{\sigma_{m,\text{sh}}^{2}}{\sigma_{m}^{2}}-\frac{{\sigma_{m,\text{sh}}^{\perp}}^{2}}{{\sigma_{m}^{\perp}}^{2}} \right)}
\end{split}
\end{equation}
The observable $\skal$ is therefore defined as
\begin{equation} \label{InvWth2}
\begin{split}
	&\skal = -
	\frac{1}{\delta_{r_{m}}^{2}} \left(\frac{\sigma_{m,\text{sh}}^{2}}{\sigma_{m}^{2}}-\frac{{\sigma_{m,\text{sh}}^{\perp}}^{2}}{{\sigma_{m}^{\perp}}^{2}} \right),
\end{split}
\end{equation}
so the effective width of $R_{\Psi_{2}}(\Delta S'')$ is $\sigma = 1/\sqrt{|\skal|}$. 
The positive (negative) $\skal$ indicates concave (convex) shape of $R_{\Psi_{2}}$, 
and zero $\skal$ indicates a flat distribution.

The measured quantity is of course 
\begin{equation} \label{skal0}
	\skal_{0} \equiv - \left( \frac{\sigma_{m,\text{sh}}^{2}}{\sigma_{m}^{2}}-\frac{{\sigma_{m,\text{sh}}^{\perp}}^{2}}{{\sigma_{m}^{\perp}}^{2}} \right)
	.
\end{equation}
Plugging Eq.~\ref{sigma2abr} into Eq.~\ref{skal0}, we have 
\begin{equation} \label{RShapeRes}
\begin{split}
	\skal_0
	=& \frac{ \shffperp^2 (1+r_{m}) + \shffpara^{2} (1-r_{m}) }{ \realperp^2 (1+r_{m}) + \realpara^{2} (1-r_{m}) } 
	-\frac{ \shffpara^2 (1+r_{m}) + \shffperp^{2} (1-r_{m}) }{ \realpara^2 (1+r_{m}) + \realperp^{2} (1-r_{m}) } \\
	=& \frac{ \realpara^{2} \shffperp^{2} - \shffpara^{2} \realperp^2 }{ (\realpara^{2} + \realperp^{2} )^2 - (\realpara^{2} - \realperp^{2})^{2} r_{m}^{2} } \times 4r_{m} \\
	\approx& 
	\frac{ \realpara^{2} \shffperp^{2} - \shffpara^{2} \realperp^2}{ (\realpara^{2} + \realperp^{2} )^2 } 
	\times 4r_{m}
	\approx - \left( \frac{\shffpara^{2}}{\realpara^{2}} - \frac{\shffperp^{2}}{\realperp^{2}} \right) r_{m}
	,
\end{split}
\end{equation}
where the approximations use $(\realpara^{2}+\realperp^{2})^{2} \gg (\realpara^{2}-\realperp^{2})^{2}$.
This measured quantity is $\skal_0=\delta_{r_{m}}^{2} \skal$ where $\delta_{r_{m}}$ is the correction factor for the resolution effect. 
This correction factor must be equal to unity when $r_{m}$ is unity, so we have
\begin{equation}\label{eq:rm}
	\delta_{r_{m}} = \sqrt{r_{m}}
	.
\end{equation}

\begin{figure}
	\includegraphics[width=1.0\linewidth]{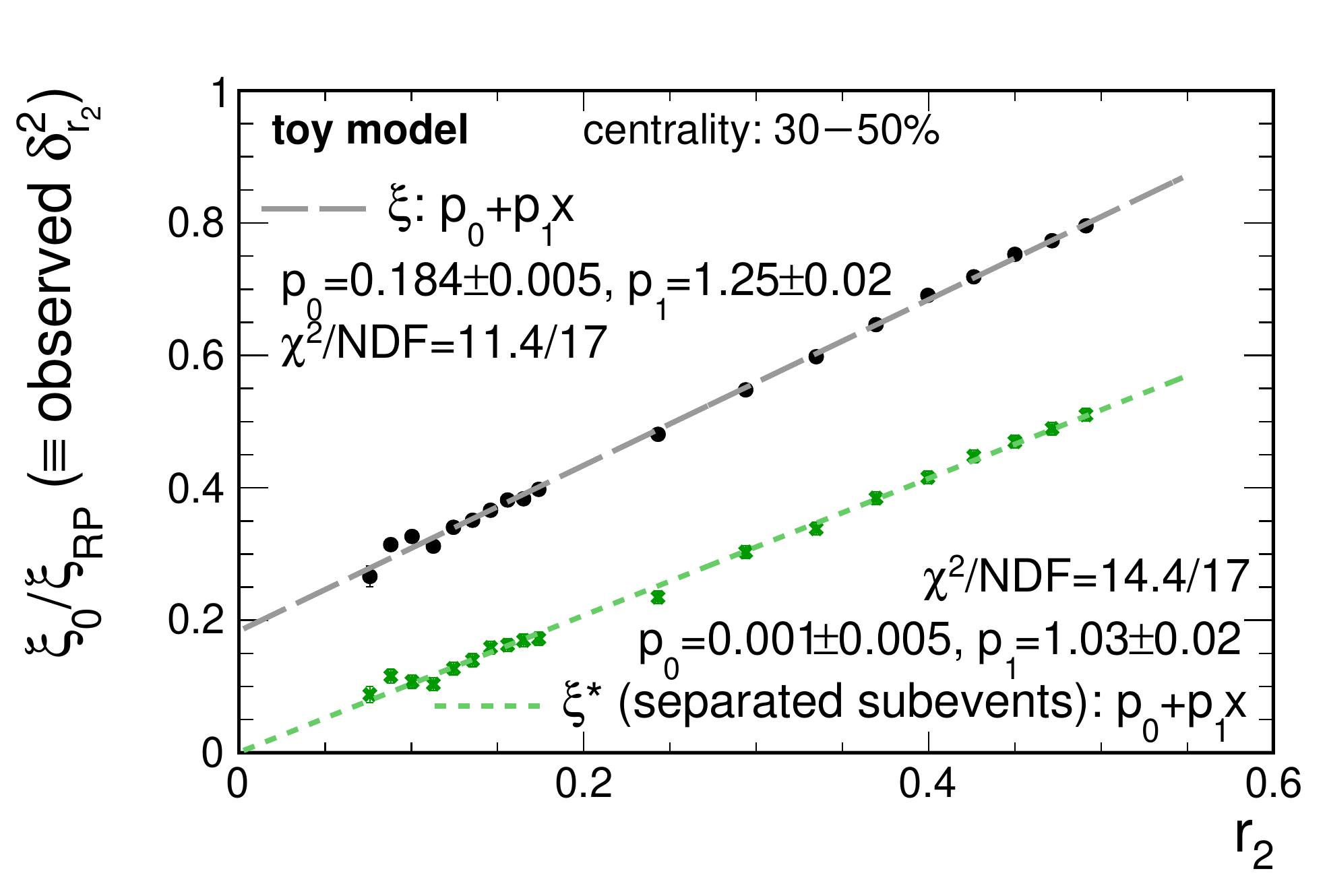}
	\caption{The $\skal_{0}/\skal_{\text{RP}}$ and $\kal_{0}/\kal_{\text{RP}}$ as functions of the EP resolution $r_{2}$ of subevents in Au+Au collisions at $\sqrt{s_{NN}} = 200 \text{ GeV}$ (no input CME signal).
	Total 4 billion toy model events in the 0--80\% centrality range are used
	(or $1.46$ billion events in the 30--50\% centrality range).
	The POI are required to have $0.35\text{ GeV/c} < p_{T} < 2.0 \text{ GeV/c}$ and $0.1< \pm \eta <1.0$,
	whereas particles for EP reconstruction are required to have 
	$0.2 \text{ GeV/c} < p_{T} < 2.0 \text{ GeV/c}$ and $0.1< \mp \eta <1.0$.
	To scan the resolution, particles for EP reconstruction are randomly kept by various probabilities 
	($100\%, 90\%, \ldots, 10\%$, and $9\%, 8\%, \ldots, 1\%$),
	while the POI are intact.
	}
	\label{ResScan}
\end{figure}

To check Eqs.~\ref{RShapeRes} and \ref{eq:rm} numerically,
a resolution scan is conducted using our toy model by randomly throwing away particles since the EP resolution is approximately proportional to the square-root of the used multiplicity.
Specifically, particles for EP reconstruction are randomly kept by various probabilities 
($100\%,\ 90\%, \ldots,\ 10\%$, and $9\%, 8\%, \ldots, 1\%$),
while the POI are intact. 
For each case, we use the measured widths
($\sigma_{2}$, $\sigma_{2,\text{sh}}$, $\sigma_{2}^{\perp}$, $\sigma_{2,\text{sh}}^{\perp}$)
to calculate the quantity $\skal_0$ by the r.h.s.~of Eq.~\ref{skal0}. 
To get the true $\skal$, we replace EP by RP since the latter is the true plane in the toy model (fixed at $\Psi_{\rm RP}=0$). The ratio of the two, $\skal_{0}/\skal_{\text{RP}}$, is the square of the correction factor, $\delta_{r_{2}}^{2}$. 
Figure~\ref{ResScan} shows the $\skal_{0}/\skal_{\text{RP}}$ 
as a function of the EP resolution $r_{2}$ of subevents 
in Au+Au collisions at $\sqrt{s_{NN}} = 200 \text{ GeV}$ (no input CME signal).
%
The $\skal_{0}/\skal_{\text{RP}}$ has linear dependence on $r_{2}$,
and the first polynomial fit gives an intercept consistent with zero
and a slope consistent with one,
as predicted by Eq.~\ref{RShapeRes}.

\subsection{The empirical EP resolution correction by STAR} \label{EPresSTAR}

In the derivation in Appendix~\ref{EPres}.\ref{EPresCalc}, we have treated 
the subevent $\Delta S^{E/W}$ separately.
If the two subevents from the same event are combined, 
as done in the STAR analysis and also studied in this work, where
the squared inverse width of $R_{\Psi_{2}}(\Delta S'')$ is referred to as $\kal$, the situation is not clear.
If the auto-correlation in Eq.~\ref{SelfCorr} was not considered, then the derivation would also hold for $\kal$ by the same math.
In the presence of auto-correlations, however, derivation of a general correction factor may not be possible because it must depend on the nature of those auto-correlations.
In the context of our toy model with the RP known, we can use the same resolution scan method described above for $\kal_{0}$ and obtain the proper resolution correction factor by $\kal_{0}/\kal_{\text{RP}}$. This is shown in Fig.~\ref{ResScan} as a function of the EP resolution $r_{2}$ of subevents. 
A linear dependence on $r_{2}$ is observed, but the intercept is nonzero. The $\kal_{0} / \kal_{\text{RP}}$ is always larger than $\skal_{0} / \skal_{\text{RP}}$, and the difference comes from those auto-correlations. See discussion in Sec.~\ref{V2Dependence} (cf.~Eq.~\ref{SelfCorr}) and 
further discussion in Appendix~\ref{EPres}.\ref{SelfKal}.

\begin{figure}
	\includegraphics[width=1.0\linewidth]{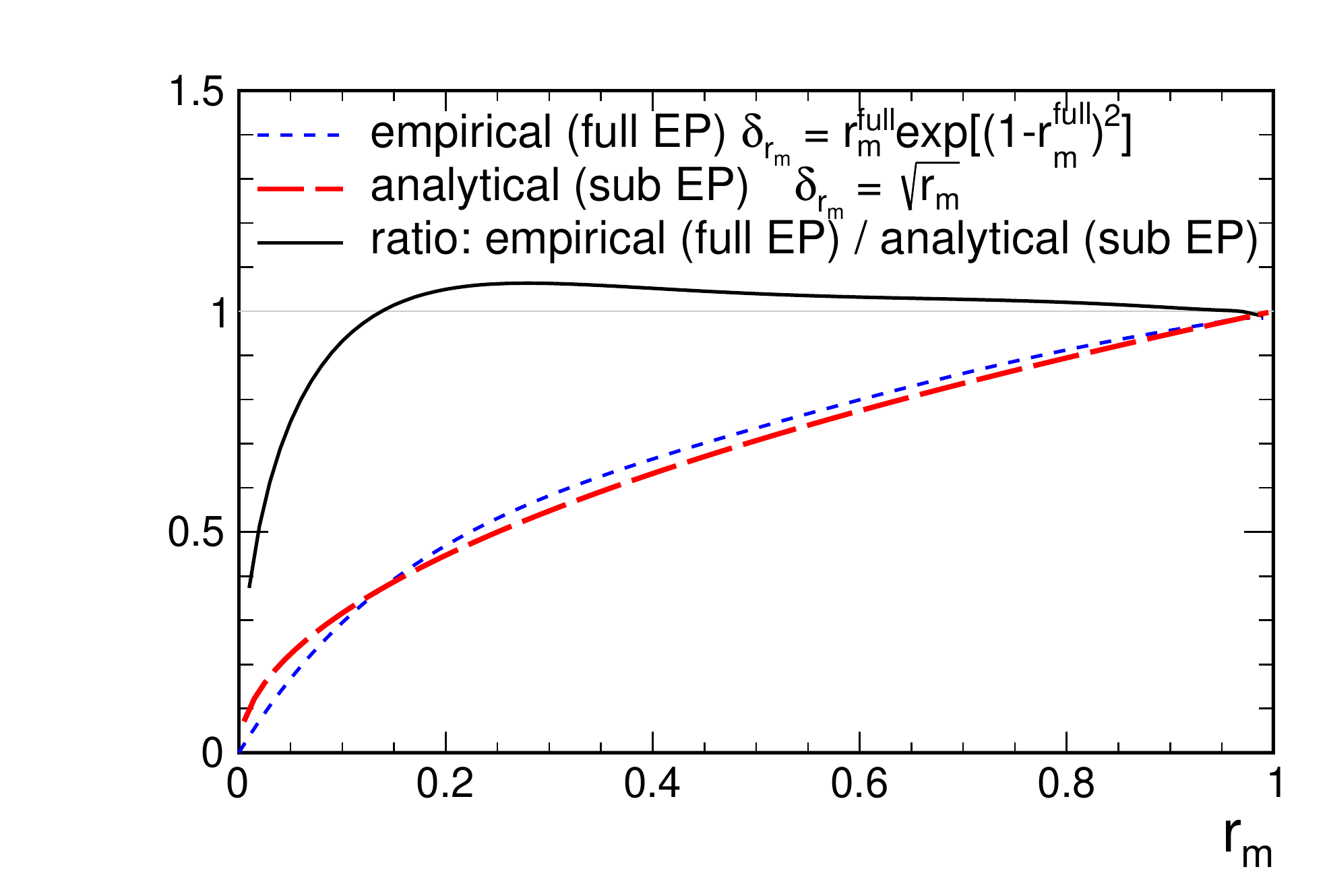}
	\caption{The analytical and empirical EP resolution correction factors plotted as functions of $r_{m}$, the EP resolution of subevents (Eq.~\ref{SubEpRes}).
	The empirical correction factor, used in the STAR paper~\cite{Magdy:2020csm}, is calculated from the EP resolution of full events ($r_{m}^{\text{full}}$), which is a monotonic function of $r_{m}$~\cite{Poskanzer:1998yz}.
	The ratio of the empirical to analytical correction factors is also shown.}
	\label{CompResFactor}
\end{figure}

The STAR paper~\cite{Magdy:2020csm} uses an empirical resolution correction factor 
$\delta_{r_{m}} = r_{m}^{\text{full}} e^{(1-r_{m}^{\text{full}})^{2}}$,
different from our analytical result of Eq.~\ref{eq:rm}.
This empirical factor uses the EP resolution of the full event, $r_{m}^{\text{full}}$,
which is a monotonic function of the EP resolution $r_{m}$ of subevents~\cite{Poskanzer:1998yz}. 
At small $r_{m}$, $r_{m}^{\text{full}}\approx \sqrt{2}r_{m}$.
The comparison between them is made in Fig.~\ref{CompResFactor}, both plotted as a function of $r_m$.
It is observed that the empirical correction factor is similar to our analytical result of $\sqrt{r_m}$.
The ratio of the two is shown by the solid curve in Fig.~\ref{CompResFactor}, indicating that the difference is less than $10\%$ 
in a wide range of resolution ($0.1 < r_{m} < 1.0$) relevant for our study.

Is the STAR empirical EP resolution correction factor correct? Figure~\ref{ResScan} suggests it is not. 
The correction factors for $\kal$ and $\skal$ are clearly different. 
The difference arises from the auto-correlations of Eq.~\ref{SelfCorr}.
The STAR empirical correction factor, which is similar to our analytical one, would be approximately correct for $\skal$, but it is incorrect for $\kal$.
In order to compare to the STAR data, we have used our analytical formula to also correct for $\kal$, which is close to the empirical factor by STAR. 
The slight difference between the two does not affect our qualitative comparisons to the STAR results.


\subsection{Auto-correlation effect in $\kal$} \label{SelfKal}

As shown in Eq.~\ref{SelfCorr}, $\kal$ contains auto-correlations,
whereas $\skal$ is free of it.
In this subsection, we investigate the effect of auto-correlations on $\kal$.
Because the $\Delta S$ distributions are even, we have $\langle \Delta S^{E} \rangle = \langle \Delta S^{W} \rangle = 0$.
The variances of $\Delta S^{E}$ and $\Delta S^{W}$ are equal to their second moments
$\text{Var}[\Delta S^{E}] = \left \langle (\Delta S^{E})^{2} \right \rangle$, 
$\text{Var}[\Delta S^{W}] = \left \langle (\Delta S^{W})^{2} \right \rangle$,
and the covariance between them is 
$\text{Cov}[\Delta S^{E}, \Delta S^{W}] = \left\langle \Delta S^{W} \Delta S^{E} \right\rangle$.
Because the two subevents are symmetric, $\Delta S^{E}$ and $\Delta S^{W}$ should have the same distribution
with the same variance $\sigma^{2} \equiv \text{Var}[\Delta S^{E}] = \text{Var}[\Delta S^{W}]$.
For convenience, we call $\rho^{2}$ the variance of full-event $\Delta S$ defined in Eq.~\ref{AveSubDS}.
Then, Eq.~\ref{SelfCorr} can be written as
\begin{equation} \label{FullWidth}
	\rho^{2} = \frac{1}{2} \sigma^{2} \left( 1 + c \right)
	,
\end{equation}
where $c$ is the correlation factor $c \equiv \text{Cov}[\Delta S^{E}, \Delta S^{W}]/\sigma^{2}$.
This applies to all four cases (real or shuffled charges, parallel or perpendicular directions).

\begin{figure}
	\includegraphics[width=1.0\linewidth]{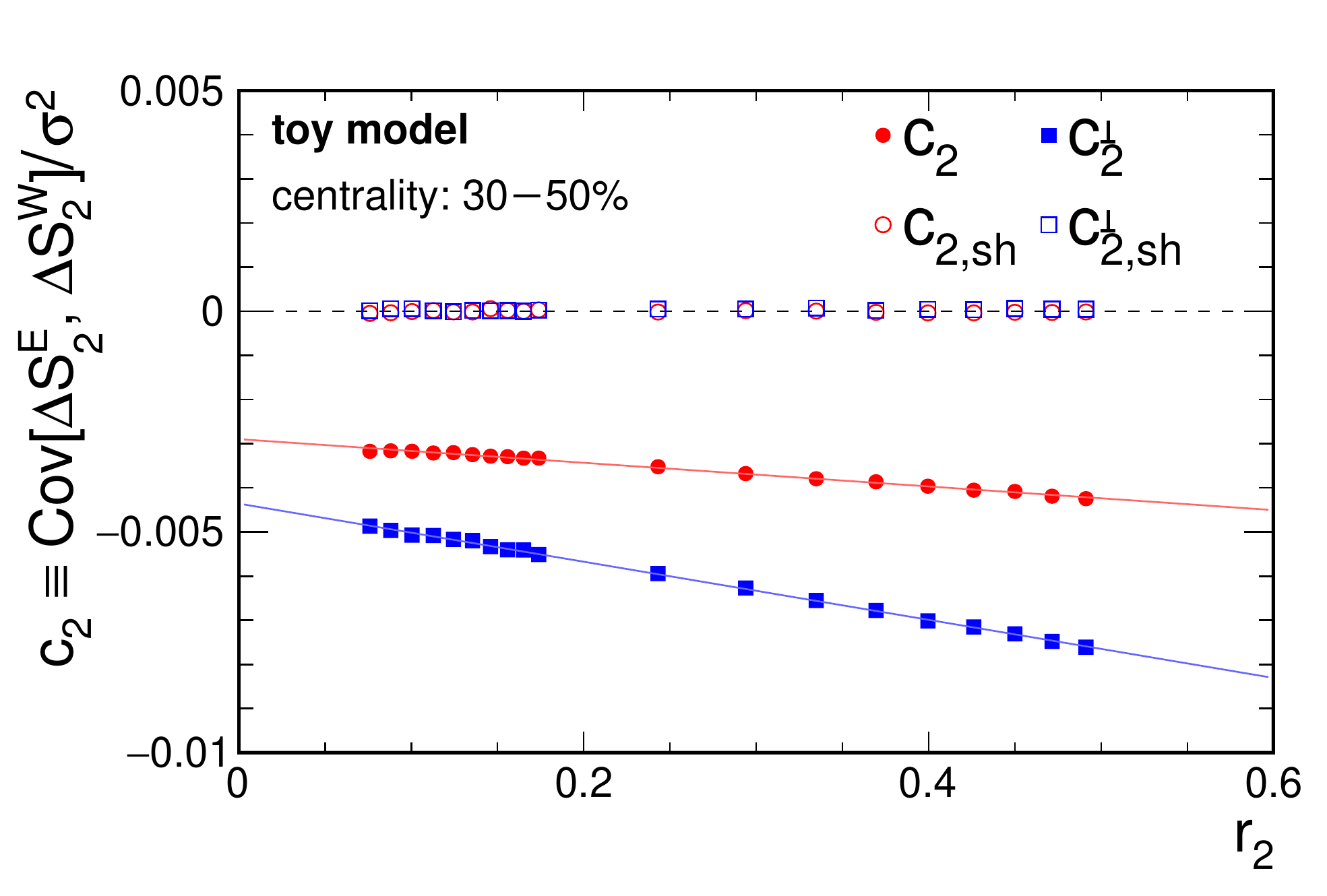}
	\caption{The correlation factors $c$ as functions of the subevent EP resolution $r_{2}$ from the same toy model simulation as Fig.~\ref{ResScan}.
	For shuffled charges, 
	the auto-correlations ($c_{2,\text{sh}}$, $c_{2,\text{sh}}^{\perp}$) vanish.
	For real charges, the auto-correlations ($c_{2}$, $c_{2}^{\perp}$) are negative, linearly dependent on $r_{2}$, and different between the two directions (parallel or perpendicular to EP). 
	If fitted by straight lines, $c_{2}$ and $c_{2}^{\perp}$ have discrepancy in their intercepts, which accounts for the nonzero intercept of $\kal_{0}/\kal_{\text{RP}}$ in Fig.~\ref{ResScan}.
	}
	\label{FctResScan}
\end{figure}

Figure~\ref{FctResScan} shows the correlation factors $c$ as functions of the subevent EP resolution $r_{2}$ 
from the same toy model simulation as Fig.~\ref{ResScan}.
For shuffled charges, the auto-correlations ($c_{2,\text{sh}}$, $c_{2,\text{sh}}^{\perp}$) vanish. 
For real charges, the auto-correlations ($c_{2}$, $c_{2}^{\perp}$) are negative and linearly dependent on $r_{2}$,
but they are different between the two directions (parallel or perpendicular to EP).
The auto-correlation factors are quite small, but they have a significant effect on $\kal$ as can be easily seen as follows.

Similar to Eq.~\ref{skal0}, we can express $\kal_{0}$ by using full-event $\Delta S$ width $\rho$
and Eq.~\ref{FullWidth}
\begin{equation} \label{kal0}
\begin{split}
	\kal_{0} \equiv& - \left( \frac{\rho_{m,\text{sh}}^{2}}{\rho_{m}^{2}}
	-\frac{{\rho_{m,\text{sh}}^{\perp}}^{2}}{{\rho_{m}^{\perp}}^{2}} \right) \\
	\approx& - \left( \frac{\sigma_{m,\text{sh}}^{2}}{\sigma_{m}^{2}} \left(1-c_{m}\right)
	-\frac{{\sigma_{m,\text{sh}}^{\perp}}^{2}}{{\sigma_{m}^{\perp}}^{2}} \left(1-c_{m}^{\perp}\right)\right) \\
	\approx & - \left( \frac{\sigma_{m,\text{sh}}^{2}}{\sigma_{m}^{2}}-\frac{{\sigma_{m,\text{sh}}^{\perp}}^{2}}{{\sigma_{m}^{\perp}}^{2}} \right) 
	+ \left( \frac{\sigma_{m,\text{sh}}^{2}}{\sigma_{m}^{2}}+\frac{{\sigma_{m,\text{sh}}^{\perp}}^{2}}{{\sigma_{m}^{\perp}}^{2}} \right) \frac{c_{m}-c_{m}^{\perp}}{2}
	,
\end{split}
\end{equation}
where $m=2$, and we have used $c_{2,\text{sh}}=c_{2,\text{sh}}^{\perp}=0$
and $|c_{2}|, |c_{2}^{\perp}| \ll 1$.
If we plug Eq.~\ref{sigma2abr} into the last line,
the first term has already been calculated in Eq.~\ref{RShapeRes},
which is $\skal_{0} \propto r_{m}$.
The quantity in the second pair of parentheses is 
\begin{equation}
\begin{split}
	& \frac{\sigma_{m,\text{sh}}^{2}}{\sigma_{m}^{2}}+\frac{{\sigma_{m,\text{sh}}^{\perp}}^{2}}{{\sigma_{m}^{\perp}}^{2}} \\
	= & 2 \cdot \frac{(\shffpara^{2}+\shffperp^{2})(\realpara^{2}+\realperp^{2})-(\shffpara^{2}-\shffperp^{2})(\realpara^{2}-\realperp^{2})r_{m}^{2}}{(\realpara^{2}+\realperp^{2})^{2}-(\realpara^{2}-\realperp^{2})^{2}r_{m}^{2}} \\
	\approx & 2 \cdot \frac{\shffpara^{2}+\shffperp^{2}}{\realpara^{2}+\realperp^{2}}
	.
\end{split}
\end{equation}
Thus, Eq.~\ref{kal0} can be written as
\begin{equation} \label{kal0cont}
\begin{split}
	\kal_{0} 
	\approx & \skal_{0} + 
	\frac{\shffpara^{2}+\shffperp^{2}}{\realpara^{2}+\realperp^{2}} \left( c_{m} - c_{m}^{\perp} \right)
	.
\end{split}
\end{equation}
The second term is the effect from auto-correlations. 
Although $(c_2-c_2^{\perp})$ is small, the coefficient in front of it is $\mathcal{O}(1)$, so it has a significant effect on the also-small quantity $\kal_{0}$.
It generally also depends on the EP resolution.
As shown in Fig.~\ref{FctResScan},
$(c_{2}-c_{2}^{\perp})$ is a first-polynomial function of $r_{2}$ with a nonzero intercept in our toy model simulation. 
Was the intercept equal to zero, we would have $\kal_{0} \propto r_{2}$ as well;
otherwise $\kal_{0}$ linearly depends on $r_{2}$ with a finite intercept, 
as shown in Fig.~\ref{ResScan}.

In general, auto-correlations should depend on the physics of the particle events. Therefore, there may not be universal resolution correction for $\kal$. 
In this work, we have used the same resolution correction factor 
for both $\kal$ and $\skal$ ($\delta_{r_{2}} = \sqrt{r_{2}}$), as stated previously,
because we want to make the comparison between $\kal$ and $\skal$, and between our analysis and Ref.~\cite{Magdy:2020csm}.

\section{Analytical form for $\skal$} \label{CalcXi}

In this appendix, we derive analytical forms of $\skal$ in the presence of $v_2$ background and CME signal.

\subsection{Background $v_{2}$ dependence} \label{BkgV2Dep}

In our previous work~\cite{Feng:2018so}, 
we derived the $v_2$ dependence of $R_{\Psi_{2}}(\Delta S)$ where $\Delta S$ was not scaled. 
In this section, we derive the $v_2$ dependence of $\skal$, 
the squared inverse width of the scaled $R_{\Psi_{2}}(\Delta S'')$ distribution, 
where the subevents are treated separately without being averaged. 
As we mentioned in Sec.~\ref{V2Dependence}, the averaging introduces auto-correlations
which make the analytical derivation inexplicable.

We only focus on the primordial pions ($\np$)
and the daughter pions from $\rho$ resonance decays ($\nr$).
The CME signal is fixed to be zero (i.e.~$a_{1} = 0$).
We assume that the number of $\pi^{+}$ and $\pi^{-}$ are the same ($\npp = \npn = \np/2$),
and denote the elliptic flow coefficients as $\vvp$ for primordial pions and $\vvr$ for $\rho$ mesons.

The analysis based on the central limit theorem (CLT)~\cite{Feng:2018so} tells us
that the widths for $\Delta S_{2}(\rp)$, $\Delta S_{2}^{\perp}(\rp)$, 
$\Delta S_{2,\text{sh}}(\rp)$, $\Delta S_{2,\text{sh}}^{\perp}(\rp)$ are
\begin{equation} \label{VarDS}
\begin{split}
	\realpara^{2} =& \frac{\nr \rk (1+\vvr) + \np(1-\vvp)}{2 (\nr + 0.5\np)^{2}} \\
	\realperp^{2} =& \frac{\nr \rk (1-\vvr) + \np(1+\vvp)}{2 (\nr + 0.5\np)^{2}} \\
	\shffpara^{2} =& \frac{2\nr(1-\vvr) + \nr\vvr\rk + \np(1-\vvp)}{2(\nr+0.5\np)^{2}} \\
	\shffperp^{2} =& \frac{2\nr(1+\vvr) - \nr\vvr\rk + \np(1+\vvp)}{2(\nr+0.5\np)^{2}}
	,
\end{split}
\end{equation}
where $\rk = \text{Var}[2\sin (\delta\phi/2)]$
is the variance of the sine value of the half decay opening angle 
($\delta\phi=\phi_{+}-\phi_{-}$, with
$\phi_{+}$ and $\phi_{-}$ being the azimuths of the $\pi^{+}$ and $\pi^{-}$ from the same $\rho$ decay).
Similar to Eq.~\ref{InvWth2}, the observable $\skal$, in which only background is present in the current case, is 
\begin{equation} \label{RPKal}
\begin{split}
	& \skal_{\text{bkgd}} = - \frac{\shffpara^{2}}{\realpara^{2}} + \frac{\shffperp^{2}}{\realperp^{2}}\\
	=& \frac{\alpha \beta \rk (4-\rk) + (\rk+2\beta-2)}
	{(\alpha \rk + 1)^{2} - (\alpha \rk \beta - 1)^{2} \vvp^{2}}
	\times 2 \alpha \vvp
	,
\end{split}
\end{equation}
where $\alpha = \nr/\np$ and $\beta = \vvr/\vvp$.
Since $\vvp \ll 1$, the second term in the denominator of Eq.~\ref{RPKal} can be safely neglected. 
We thus have
\begin{equation} \label{CalcKalV2}
\begin{split}
	\skal_{\text{bkgd}} \approx 
	\frac{\alpha \beta \rk (4 - \rk) + (\rk+2\beta-2)}
	{(\alpha \rk + 1)^{2}}
	\times 2 \alpha \vvp 
	.
\end{split}
\end{equation}
We can see that, for pure background, 
$\skal$ is approximately proportional to the background $v_{2}$.

In our toy model simulations, 
the multiplicity ratio of $\rho$ to primordial pions is $0.085$. 
To have a rough estimate, we can take $\alpha=0.085$. 
The $\vvr$ is parameterized taking into account the NCQ scaling at high $p_{T}$ 
and the hydrodynamics mass ordering of $v_{2}$ at low $p_{T}$;
we find $\beta \approx 1.1$. 
From our previous study, we found the RMS of $2\sin(\delta\phi/2)$ is $1.36$~\cite{Feng:2018so}, 
thus $\rk=1.36^2=1.85$. 
With these eatimates of $\alpha$, $\beta$, $\rk$, we have 
\begin{equation}
	\skal_{\text{bkgd}} \approx 0.31 \vvp
	.
\end{equation}
This is about a factor of 3 larger compared to our toy model simulation in Sec.~\ref{V2Dependence} (cf.~Eq.~\ref{TMv2Dep}b).
In our derivation here, we have simply assumed that all $\rho$ decay daughters are included in the POI's. 
In the toy model using subevents, 
only a fraction of the $\rho$ resonances have both daughters in the subevent acceptance. 
This would significantly reduce the coefficient in the toy model compared to the derivation in Eq.~\ref{CalcKalV2}. 
Other simplifying assumptions, such as neglecting correlations arising from $p_{T}$ dependence of $v_{2}$ and decay kinematics, may also contribute to the numerical difference. 
However, the qualitative features in the results from the analytical derivation, 
namely the proportionality to $v_{2}$ and the $\mult$ independence, are robust and provide useful insights.


\subsection{Signal $a_{1}$ dependence} \label{SigA1Dep}


In this section, we derive analytically the dependence of $\skal$ on the CME signal strength, $a_{1}$.
The primordial particle $a_{1}$ in Eq.~\ref{PrimoDist} is nonzero for each event,
but it can be either positive or negative for different events,
so the event average of $a_{1}$ is still zero.
On one hand, the positive and negative charges in the same event always have opposite $a_{1}$,
so shuffling the charges removes the signal contribution.
On the other hand, the signal only contributes to the charge separation in the $y$-direction,
so the $x$-projection is not affected.
Thus, only the distribution of $\Delta S_{m}$ has dependence on $a_{1}$.
In this section, we will only focus on the second-order $\Delta S_{2}$ with respect to RP, namely
\begin{equation}
	\Delta S = \frac{1}{n^{+}} \sum_{i}^{n^{+}} \sin(\phi_{i}^{+}-\rp)
	- \frac{1}{n^{-}} \sum_{i}^{n^{-}} \sin(\phi_{i}^{-}-\rp)
	,
\end{equation}
where we assume $n^{+} = n^{-} = 0.5 n_{\pi} + n_{\rho}$ as mentioned in Appendix~\ref{CalcXi}.\ref{BkgV2Dep}.

We first fix $a_{1}$ and get the conditional expectation and variance of $\Delta S$.
It is straightforward that the average of $\pm \sin(\phi^{\pm} - \rp )$ 
over all primordial particles among all events is 
\begin{equation}
	\text{E} \left[ \left. \pm \sin(\phi^{\pm}-\rp) \right| a_{1} \right] = a_{1}
	,
\end{equation}
and the conditional variance is
\begin{equation}
\begin{split}
	& \text{Var} \left[ \left.\pm \sin(\phi^{\pm} - \rp ) \right| a_{1} \right] \\
	= &\langle \sin^{2} (\phi^{\pm} - \rp ) \rangle - \langle \pm \sin(\phi^{\pm}-\rp) \rangle^{2} \\
	= & \frac{1 - \vvp - 2 a_{1}^{2}}{2}
	.
\end{split}
\end{equation}
Thus, we can get the conditional variance of $\Delta S$ by substituting $\vvp$ by $\vvp + 2 a_{1}^{2}$
for $\realpara^{2}$ in the first line of Eq.~\ref{VarDS}, namely
\begin{equation}
	\text{Var}\left[ \left. \Delta S \right| a_{1} \right] =
	\frac{\nr \rk (1+\vvr) + \np(1-\vvp - 2 a_{1}^{2})}{2 (\nr + 0.5\np)^{2}}
	.
\end{equation}
We can also get the conditional expectation of $\Delta S$
\begin{equation}
	\text{E} \left[ \left. \Delta S \right| a_{1} \right] = \frac{ \np}{0.5\np + \nr} a_{1}
	.
\end{equation}

Now with varying $a_{1}$ from event to event, 
since the topologic charge fluctuation is totally random among events,
$a_{1}$ is a symmetric distribution about $0$,
so $\text{Var} \left[ a_{1} \right] = \langle a_{1}^{2} \rangle$.
Thus, the total variance can be calculated
\begin{equation}
\begin{split}
	& \realpara^{2}(\langle a_{1}^{2} \rangle) = \text{Var} \left[ \Delta S \right] \\
	=& \text{E} \left[ \text{Var}\left[ \left. \Delta S \right| a_{1} \right] \right]
	+ \text{Var} \left[ \text{E} \left[ \left. \Delta S \right| a_{1} \right] \right] \\
	\approx & \frac{\nr \rk (1+\vvr) + \np(1-\vvp - 2 \langle a_{1}^{2} \rangle)}{2 (\nr + 0.5\np)^{2}} \\
	&+ \left(\frac{ \np}{0.5\np + \nr}\right)^{2} \langle a_{1}^{2} \rangle \\
	\approx & \realpara^{2}(0) + \left(\frac{ \np}{0.5\np + \nr}\right)^{2} \langle a_{1}^{2} \rangle
	,
\end{split}
\end{equation}
where we have assumed $\vvp \gg a_{1}^{2}$ so the latter is dropped from the first term, 
which is then simply given by Eq.~\ref{VarDS} without the $a_{1}$ signal.

Again, for convenience, we write all $\Delta S$ distributions as Gaussians,
Eq.~\ref{RPKal} would be modified, with finite $a_{1}$, into
\begin{equation} \label{KalTot}
\begin{split}
	& \skal ( \langle a_{1}^{2} \rangle ) 
	= - \frac{\shffpara^{2}}{\realpara^{2}( \langle a_{1}^{2} \rangle )} + \frac{\shffperp^{2}}{\realperp^{2}}\\
	=& - \frac{\shffpara^{2}}{\realpara^{2}(0)} \left( 1 + \left(\frac{ \np}{0.5\np + \nr}\right)^{2}  \frac{\langle a_{1}^{2}\rangle}{\realpara^{2}(0)} \right)^{-1}
	+ \frac{\shffperp^{2}}{\realperp^{2}}\\
	\approx & - \frac{\shffpara^{2}}{\realpara^{2}(0)} + \frac{\shffperp^{2}}{\realperp^{2}}
	+ \left(\frac{ \np}{0.5\np + \nr}\right)^{2} \frac{\shffpara^{2}  }{\realpara^{4}(0)} \langle a_{1}^{2} \rangle \\
	\approx & \skal_{\text{bkgd}} + 2 \mult ( 1 + \vvp - 2 \alpha K ) \langle a_{1}^{2}  \rangle
	,
\end{split}
\end{equation}
where the first term, $\skal_{\text{bkgd}}$, is that given by Eq.~\ref{CalcKalV2} without $a_{1}$ signal.
The first approximation comes from the fact that $\realpara^{2}(0) \gg  \langle a_{1}^{2} \rangle$,
as $\realpara^{2}(0) \sim 1/\np \sim 10^{-2}$ and $\langle a_{1}^{2} \rangle \sim 10^{-4}$.
We simply take the number of POI's as $\mult \approx \np + 2\nr$.
We can see from Eq.~\ref{KalTot} that the background and the CME are approximately decoupled in $\skal$, 
and the CME signal $\skal(\langle a_{1}^{2} \rangle)$ has linear dependence on $\langle a_{1}^{2} \rangle$.

With the aforementioned values for $\alpha$ and $K$, we can estimate the signal contribution to be
\begin{equation}
	\skal_{\text{CME}} \approx 1.49 \langle a_{1}^{2} \rangle \mult
	.
\end{equation}
This is close to the toy model simulation result in Eq.~\ref{TMa1v2Dep}b.
The $S/B$ of $\skal(\langle a_{1}^{2} \rangle)$ can be estimated as
\begin{equation} \label{CalcKalBS}
\begin{split}
	S/B = \frac{1.49 \langle a_{1}^{2} \rangle \mult }{ 0.34 \vvp }
	= 4.8 \langle a_{1}^{2} \rangle \mult / \vvp
	,
\end{split}
\end{equation}
which is about a factor of 3 smaller than the toy model result of Eq.~\ref{TMa1v2SB}b,
mainly inherited from the discrepancy in $\skal_{\text{bkgd}}$ estimation in Appendix~\ref{CalcXi}.\ref{BkgV2Dep}.


\section{Analytical form for $\Delta\gamma$} \label{CalcDg}

For completeness, we can easily obtain $\Delta\gamma$ from Eqs.~\ref{subgamma}-\ref{GammaDFv2}.
\begin{equation} \label{CalcGammaDF}
\begin{split}
	\Delta\gamma \approx &
	\frac{ \nr }{ (\np/2 + \nr)^{2} } \vvr D
	+ 2 \left( \frac{ \np/2 }{\nr + \np/2} \right)^{2} \langle a_{1}^{2} \rangle \\
	= & \frac{4\alpha D \beta}{1+2\alpha} \frac{\vvp}{\mult} + \frac{2}{(1 + 2\alpha)^{2}} \langle a_{1}^{2} \rangle
	,
\end{split}
\end{equation}
where $D = \langle \cos (\phi_{a} + \phi_{b} - 2\phi_{\text{reso}}) \rangle$ as shown in Eq.~\ref{GammaDFv2}.
Taking $D \sim 0.65$~\cite{Wang:2016iov} and the aforementioned $\alpha$ and $\beta$ values, 
we obtain $\Delta\gamma$ for our toy model setting in this work as
\begin{equation}
	\Delta\gamma \approx 0.21 \vvp / \mult + 1.46 \langle a_{1}^{2} \rangle
	.
\end{equation}
The $S/B$ for $\Delta\gamma$ is then
\begin{equation}
\begin{split}
	\frac{ \np^{2} }{2\nr} \frac{\langle a_{1}^{2} \rangle}{ \vvr D}
	=& \frac{1}{2 \alpha(1+2\alpha) \beta D} \mult \langle a_{1}^{2} \rangle / \vvp \\
	\approx & 7.0 \mult \langle a_{1}^{2} \rangle / \vvp
	.
\end{split}
\end{equation}

The proportionality coefficient on $a_{1}^{2}$ is close to 
that obtained from our toy model simulation in Eq.~\ref{TMa1v2Dep}c.
The coefficient on the $v_{2}$ background is about a factor of 2 larger than that from the toy model simulation.
This arises from 
similar reasons responsible for the discrepancy in $\skal_{\text{bkgd}}$ between the analytical estimate and the toy model simulation. 
Namely, not all $\rho$ resonances have both decay daughters in the subevent acceptance, 
and correlations exist among various quantities because of their dependences on $p_{T}$.
Note that those effects appear to yield a larger discrepancy in $\skal$ than in $\Delta\gamma$. 
As a result, the $S/B$ seems better for $\skal$ than $\Delta\gamma$ in our toy model simulation, 
and it is reversed in the analytical results. 
This quantitative feature likely depends on the details of the model implementation, 
such as the types of resonances included and their abundances. 

\bibliography{./ref}


\end{document}